\documentclass[10pt,a4paper]{article}

\usepackage{amssymb}
\usepackage{amsfonts}
\usepackage{mathrsfs} 
\usepackage{latexsym}
\usepackage{amsmath}
\usepackage{graphics}
\usepackage{graphicx}
\usepackage{hhline}
\usepackage{euscript}
\usepackage{rotating}
\usepackage{setspace}

\DeclareSymbolFont{NumBold}{U}{bbold}{m}{n}
\DeclareSymbolFontAlphabet{\mathNumbb}{NumBold}
\DeclareMathSymbol{\Id}{\mathord}{NumBold}{`1}
\newcommand{\mbf}[1]{\mbox{\boldmath{$#1$}}}

\newcommand{\e}{\mathop{\mathrm{e}}\nolimits}
\newcommand{\Tr}{\mathop{\mathrm{Tr}}\nolimits}
\newcommand{\Ref}[1]{(\ref{#1})}

\newcommand{\dd}{\mathrm{d}}

\newcommand{\alg}{\mathfrak}
\newcommand{\diag}{\mathop{\mathrm{diag}}\nolimits}
\newcommand{\pd}{\partial}
\newcommand{\Real}{\mathbb{R}}
\newcommand{\Compl}{\mathbb{C}}
\newcommand{\mr}{\mathrm}
\newcommand{\tr}{\mr{T}}

\newcommand{\vpint}{\displaystyle -\hskip -1.1em \int}
\renewcommand{\k}{\mbf{k}}
\newcommand{\Dirac}{\nabla}
\newcommand{\Sym}{\mathop{\mathrm{Sym}}\nolimits}
\newcommand{\Ant}{\wedge}
\newcommand{\Tor}{\mathbb{T}}
\newcommand{\Sphere}{\mathbb{S}}
\newcommand{\Mod}{\mathfrak{M}}
\newcommand{\Hom}{\mathop{\mathrm{Hom}}\nolimits}
\newcommand{\Prep}{\mathcal{F}}
\newcommand{\Ch}{\mathop{\mathrm{Ch}}\nolimits}
\newcommand{\Td}{\mathop{\mathrm{Td}}\nolimits}
\newcommand{\Ind}{\mathrm{Ind}}
\newcommand{\Heavi}{\mathscr{H}}

\newcommand{\sn}{\mathop{\mathrm{sn}}\nolimits}
\newcommand{\logmod}[1]{\ln\left| #1 \right| }
\newcommand{\logmodL}[1]{\logmod{\frac{#1}{\Lambda}}}
\newcommand{\bA}{\mathbb{A}}
\newcommand{\bJ}{\mathbb{J}}
\newcommand{\A}{\mathcal{A}}
\newcommand{\B}{\mathcal{B}}

\newcommand{\E}{\mathcal{E}}

\newcommand{\G}{\mathcal{G}}

\renewcommand{\L}{\mathcal{L}}
\newcommand{\M}{\mathcal{M}}
\newcommand{\N}{\mathcal{N}}

\renewcommand{\P}{\mathcal{P}}

\newcommand{\R}{\mathcal{R}}
\renewcommand{\S}{\mathcal{S}}

\newcommand{\V}{\mathcal{V}}
\newcommand{\W}{\mathcal{W}}
\renewcommand{\a}{\alpha}
\renewcommand{\b}{\beta}
\newcommand{\g}{\gamma}
\renewcommand{\d}{\delta}
\newcommand{\eps}{\varepsilon}
\newcommand{\ep}{\epsilon}
\renewcommand{\o}{\omega}
\renewcommand{\r}{\rho}
\renewcommand{\c}{\chi}
\newcommand{\f}{\phi}
\newcommand{\vf}{\varphi}
\renewcommand{\l}{\lambda}
\newcommand{\s}{\sigma}
\renewcommand{\t}{\tau}
\newcommand{\vth}{\vartheta}
\newenvironment{remark}{\noindent\underline{Remark.}}{$\Box$}
\newcommand{\microsection}[1]{\noindent\underline{#1}}

\begin{document}

\setstretch{1.5}

%%% Title Page %%%
\makeatletter
\begin{titlepage}
\begin{flushright}
{hep-th/yymmnn}\\
{IHES/P/04/38}
\end{flushright}
\begin{center}
\vskip .5in
{\LARGE\bf Saddle point equations in Seiberg-Witten theory}\\
\vskip 2cm
{\bf Sergey Shadchin}
\end{center}
\bigskip
\bigskip
\centerline{\em Institut des Hautes Etudes Scientifiques}
\centerline{\em 35 route de Chartres, 91440 Bures-sur-Yvette, FRANCE}
\centerline{email: {{\tt chadtchi@ihes.fr}}}
\bigskip
\abstract{\noindent $\N=2$ supersymmetric Yang-Mills theories for all classical gauge groups, that is, for $SU(N)$, $SO(N)$, and $Sp(N)$ is considered. The equations which define the Seiberg-Witten curve are proposed. In some cases they are solved. It is shown that for (almost) all models allowed by the asymptotic freedom the 1-instanton corrections which follows from these equations agree with the direct computations and with known results.}
\end{titlepage}
\makeatother

%%% Table of Contents %%%

\tableofcontents

%%% Introduction %%%

\section{Introduction}

After its appearance the Seiberg-Witten solution for $\N = 2$ supersymmetric Yang-Mills theory for the gauge group $SU(2)$ \cite{SeibergWitten} was generalized in both directions: to the other classical groups and to various matter content \cite{SimpleSing,VacuumStruct,QuantModSpace,CurveForSO(2r+1),HyEl,SeibergWittenII,NewCurves,SYMandIntegr}. 

Till recently while generalizing one established the expression for the algebraic curve and the meromorphic differential from the first principles and then computed the instanton corrections to the (up to two-derivatives and four-fermions pieces) effective action. This action can be expressed with the help of unique holomorphic function $\Prep(a)$, referred as prepotential. With the help of extended superfield formalism the Lagrangian for the effective theory can be written as follows:
$$
\frac{1}{8\pi h^\vee} \Im \int \dd^4 \theta \Prep(\Psi).
$$
The classical prepotential, which provides the microscopic action, is
\begin{equation}
\label{PrepClass}
\Prep_{\mr{class}}(a) = \pi i \t \langle a,a \rangle,
\end{equation}
where $\displaystyle \t = \frac{4\pi i}{g^2} + \frac{\Theta}{2\pi}$. Note that we use the normalization of the prepotential which differs from some other sources by the factor $2\pi i$. Let us write here, for further references, the perturbative contribution to the prepotential, which comes from the 1-loop computations:
\begin{equation}
\label{PrepPert}
\begin{aligned}
\Prep_{\mr{pert}}(a,m) &= - \frac{1}{2} {\langle a,\a\rangle}^2 \left( \ln\left|\frac{\langle a,\a \rangle }{\Lambda}\right| - \frac{3}{2}\right) \\
&+ \frac{1}{2} \sum_{r \in \mr{reps}} \sum_{\l \in \mbf{w}_r}{\left(\langle a,\l \rangle + m_r\right)}^2 \left( \ln \left|\frac{\langle a,\l \rangle + m_r}{\Lambda}\right| - \frac{3}{2}\right)
\end{aligned}
\end{equation}
where $\R$ is the root system of the gauge group and $\mbf{w}_r$ is the set of all  weights of the representation $r$. The first sum in the second line is taken over all the matter representation contained in the model.

In \cite{SWfromInst,SWandRP} a powerful technique was proposed to follow this way in the opposite direction: to compute first the instanton corrections and to extract from them the Seiberg-Witten geometry and the analytical properties of the prepotential.

In \cite{ABCD}  the solution of $\N = 2$ supersymmetric Yang-Mills theory for the classical groups other that $SU(N)$ using the method proposed in \cite{SWfromInst,SWandRP} was obtained. This method consists of the reducing functional integral expression for the vacuum expectation of an observable (in fact, this observable equals to 1, hence we actually compute the partition function as it defined in statistical physics) to the finite dimensional moduli space of zero modes of the theory. That is, to the instanton moduli space, the moduli space of the solutions of the self-dual equation
\begin{equation}
\label{SDE}
F_{\mu\nu} - \star F_{\mu\nu} = 0
\end{equation}
with the fixed value of the instanton number, defined as follows:
\begin{equation}
\label{Inst}
k = -\frac{1}{16\pi h^\vee} \int \Tr_{\mr{adj}} F \wedge F. 
\end{equation}
Notation $\Tr_{\mr{adj}}$ means that the trace is taken over the adjoint representation.

In this paper we continue to investigate the possibility to solve the $\N = 2$ supersymmetric Yang-Mills theory with various matter content (limited, of cause, by the asymptotic freedom condition).

Roughly speaking our task can be split into two parts. First part consists of the writing the expression for the finite dimensional integral to which vacuum expectation in question can be reduced. To accomplish this task in \cite{SWfromInst,ABCD} the explicit construction for the instanton moduli space was used. Already for the pure gauge theory its construction (the famous ADHM construction of instantons, \cite{ADHM}) is rather nontrivial (see for example \cite{MultInstMeasure,MultiInstCalc,MultiInstCalcII,ItoSasakuraSU,ItoSasakura,ItoSasakuraSUSYQCD}). In the presence of matter it becomes even more complicated. 

Fortunately there is another method which lets to skip the explicit description of the moduli space and to directly write the required integral. This method uses some algebraic facts about the universal bundle over the instanton moduli space. It will be explained in section \ref{UniversalBundle}. Using this method we will obtain the prepotential as a formal series over the dynamically generated scale.

The second part of the task is to extract the Seiberg-Witten geometry from obtained expressions. To do this we will use the technique proposed in \cite{SWandRP}. It is based on the fact that in the limit of large  instanton number the integral can be estimated by means of the saddle point approximation. This approximation can be effectively described by the Seiberg-Witten data --- the curve and the differential. One may wonder why the prescription obtained in this limit will provide the exact solution even in the region of finite $k$, where the saddle point approximation  certainly will not work. The answer is that the real reason why the Seiberg-Witten prescription works is the holomorphicity of the prepotential, pointed out in \cite{SeibergWitten}, whereas the saddle point approximation just makes it evident and easy to extract.

The paper is organized as follows: in section \ref{I} we describe a method to write the formula for the instanton corrections. In section \ref{II} we reduce the problem of the instanton correction computations to the problem of minimizing a functional. And finally in section \ref{III} we solve the saddle point equations for some models. Using relations between the saddle point equation for different models we establish the same relations between the prepotentials for these models and finally we find the {\em hyperelliptic approximation} for the Seiberg-Witten curves for all the models. This allows us to compute the 1-instanton corrections which comes from the algebraic curve computation and compare it with the direct computation. In each case perfect agreement between results of the two approaches is found. 

%%% Toward the formal expression %%%

\section{Episode I: Toward the formal expression}
\label{I}

The RG-flow for the coupling constant $\t$ is given by 
$$
\t(\Lambda_1) = \t(\Lambda_2) + \frac{\b}{2\pi i} \ln \frac{\Lambda_1}{\Lambda_2},
$$ 
where the $\b$-function leading term is defined by the coefficient
\begin{equation}
\label{ass}
\beta = \zeta \left(\ell_{\mr{adj}} - \sum_{r \in\mr{reps}} \ell_r \right),
\end{equation}
$\ell_r$ being the Dynkin index of the representation $r$ of the gauge group $G$. The Dynkin indices are listed in Table \ref{GroupTh}. For all groups $\ell_{\mr{fund}} = 1$. $\zeta = 1$ for the unitary group and $\zeta = 2$ for orthogonal and symplectic group.

Let us choose the energy scale in such a way, that the RG-flow equation becomes $\t(\Lambda) = \t_0 + \dfrac{\b}{2\pi i} \ln \Lambda$. Introduce the instanton counting parameter
$$q = \e^{2\pi i \t} = \e^{2\pi i \t_0} \Lambda^\b.$$
In this section we obtain the formal expression for the partition function 
\begin{equation}
\label{Z}
Z(a,m,\eps_1,\eps_2,\Lambda) = Z^{\mr{pert}}(a,m) \times \sum_{k=0}^\infty q^k Z_k(a,m,\eps) = \exp{\frac{1}{\eps_1\eps_2}\Prep(a,m,\eps)},
\end{equation}
where $Z^{\mr{pert}}(a,m)$ is the perturbative contribution to the partition function,  the prepotential of $\N = 2$ super Yang-Mills theory is given by $\lim_{\eps_1,\eps_2\to 0} \Prep(a,m,\eps) \equiv \Prep(a,m)$.

\begin{remark}
When $\b\neq 0$ we can completely neglect $\t_0$ and in this case we have $q \mapsto \Lambda^\b$. For the conformal theories, that is, for the theories where $\beta = 0$, we have $q = \e^{2\pi i \t_0}$. 
\end{remark}

Each summand in \Ref{Z} comes from the localization of the initial path integral on the $k$-instanton solution. This localization can be performed thanks to the Duistermaat-Heckman (DH) formula \cite{Duistermaat-Heckman}. 

Recall the main steps of this calculation. Let $\M$ be a $2n$-dimensional symplectic manifold with the symplectic form $\o$. Let a torus $\Tor$ act symplectically, and suppose that its action can be described by a Hamiltonian (momentum) map $\mu:\M \to \alg{t}^\ast$, $\alg{t} = \mbox{Lie}(\Tor)$. The choice of $x \in \alg{t}$ defines the Hamiltonian $H = \langle \mu, x \rangle$ and the action. Let $p_f \in M$ be a fixed point this action and $w_\a(p_f) \in \alg{t}^\ast$ a weight of this action on the tangent space to $p_f$. Then the DH formula states
\begin{equation}
\label{DH}
\int_\M \frac{\o^n}{n!} \e^{-\langle \mu, x \rangle} = \sum_{p_f} \frac{\e^{-\langle\mu(p_f),x\rangle}}{\prod_{\a} \langle w_\a(p_f), x\rangle}.
\end{equation}
In that follows we will use the shorthand notation $\langle w_\a(p), x \rangle \equiv w_\a$.

\begin{remark}
When we deal with the supermanifolds, which contain supercoordinates, the DH formula should be modified as follows: $\prod_\a w_\a \mapsto \prod_\a {w_\a}^{\ep_\a}$ where $\ep_\a = \pm 1$ depends on the statistics of coordinate it comes from.
\end{remark}

It turns out to be easier to compute first the character of the torus element $q$:
$$
\Ind_q \equiv \sum_\a \ep_\a e^{w_\a}.
$$
This can be done with the help of the equivariant analogue of the Atiyah-Singer index theorem. It worth noting that when $\Ind_q$ is derived equivariantely, the sings $\ep_\a$ comes from the alternating summation over cohomologies, and not from bosonic-fermionic statistics.  

Once we have $\Ind_q$, the passage to the DH formula can be done with the help of the following transformation (which can be seen as a proper time regularization, see section \ref{SU(N)PureYM}):
\begin{equation}
\label{TrWeights}
\sum_{\a} \ep_\a \e^{w_\a} \mapsto \prod_\a {w_\a}^{\ep_\a}.
\end{equation}

%%% ADHM data %%%

\subsection{ADHM construction}

The moduli space of solutions of the self-duality equation \Ref{SDE} with the fixed value of the instanton number $k$ \Ref{Inst} can be described as follows \cite{ADHM,SelfDualSolution,InstAndRec,CalcManyInst}. First of all we introduce two vector spaces $\V$ and $\W$. The space $\V$ is the space of the fundamental representation of the dual group $G_D$, $\W$ -- of the gauge group $G$. Their dimensions together with the sizes of the matrices introduces below  can be found in the Table \ref{ADHMTable}. Also we introduce the spinor coordinate and the complex structure on $\Real^4 \simeq \Compl^2$ as follows:
$$x_{\a\dot{\a}} = x^\mu \sigma_{\mu,\a\dot{\a}} = \left(
\begin{array}{cc}
x^0 -i x^3 &  -i x^1 - x^2 \\
- i x^1 + x^2 & x^0 + i x^3
\end{array}
\right) = \left(
\begin{array}{cc}
z_1 & -z_2^\ast \\
z_2 & z_1^\ast
\end{array}
\right).
$$

Define a matrix $\Delta_{\dot{\a}}$ linearly depending on $x_{\a\dot{\a}}$:
$$
\Delta_{\dot{\a}}(x) = \A_{\dot{\a}} + \B^\a x_{\a\dot{\a}}.
$$
The matrix $\Delta_{\dot{\a}}$ obeys the so called factorization condition
\begin{equation}
\label{fac}
\Delta^{\dag\dot{\a}} \Delta_{\dot{\b}} = \delta_{\dot{\b}}^{\dot{\a}} \R^{-1}
\end{equation}
where $\R$ is an invertible matrix.

To be specific we consider the case of $G = U(N)$. For others groups see remark in the end of section. 

In order to recover the connection we require a matrix $v$  satisfying the following conditions:
\begin{equation}
\label{vCond}
\begin{aligned}
v^\dag v &= \Id_\W & &\mbox{and} & v^\dag \Delta_{\dot{\a}} = 0. 
\end{aligned}
\end{equation}

Once we get such a matrix we can write the connection as follows:
$$
A_\mu = v^\dag \pd_\mu v.
$$

Moreover using this data we can obtain all the (normalizable) solutions of the Weyl equation in the instanton background. We introduce the Dirac operator $\Dirac_{\a\dot{\a}} = (\pd_\mu + A_\mu)\sigma^\mu_{\a\dot{\a}}$ and for solutions of the Weyl equation
\begin{equation}
\label{WEQ}
\Dirac^{\dag \dot{\a}\a}\psi_\a = 0
\end{equation}
we have the following expression \cite{OsbornDirac}:
\begin{equation}
\label{SolWeyl}
\psi^\a = v^\dag \B^\a \R.
\end{equation}
These solutions are normalized as follows:
$$
\int \dd^4 \psi^\dag_\a \psi^\a = \pi^2 \Id_\V.
$$

Without loss of generality we can assume that the matrix $\B^\a$ has the following form
$$
\B = (\B^1, \B^2) = \left(
\begin{array}{c}
0 \\
\Id_\V \otimes \Id_2
\end{array}
\right).
$$
Then the relevant data is contained in the matrices $\A_{\dot{\a}}$ and $v$ which can be represented as follows
$$
\begin{aligned}
\A = (\A_{\dot{1}},\A_{\dot{2}}) &= \left(
\begin{array}{c}
S_{\dot{1}}, S_{\dot{2}} \\
X^\mu \otimes \sigma_{\mu}
\end{array}
\right), & v &= \left(
\begin{array}{c}
T \\
Q_\a
\end{array}
\right).
\end{aligned}
$$

The factorization condition \Ref{fac} can be rewritten for 
$$(B_1,B_2,I,J) = (X^0 - i X^3,X^2 - i X^1,{S_{\dot{2}}}^\dag,S_{\dot{1}})$$
as follows (the ADHM equations):
\begin{equation}
\label{ADHM}
\begin{aligned}
\mu_\Compl &= {[}B_1,B_2{]} + IJ = 0 \\
\mu_\Real &= [B_1,{B_1}^\dag] + [B_2,{B_2}^\dag] + II^\dag - J^\dag J &= 0.
\end{aligned}
\end{equation}

The description presented above is a little redundant. Namely one can see that the following transformation does not change the connection
$$
\begin{aligned}
S_{\dot{\a}} &\mapsto S_{\dot{\a}}' = g S_{\dot{\a}} g_D^{-1}, & X^\mu &\mapsto {X^\mu}' = g_D X^\mu g_D^{-1},\\
T &\mapsto T' = gT, & Q_\a &\mapsto Q'_\a = g_D Q_\a 
\end{aligned}
$$
where $g \in G$, $g_D \in G_D$ (see Table \ref{ADHMTable}).

The properties of the introduces matrices with respect to the space-time rotations are the following: $S_{\dot{\a}}$ is a dotted spinor, $X^\mu$ is a vector. For the second equation of \Ref{vCond} to be Lorentz invariant the matrix $T$ should be a scalar and $Q_\a$ should be an undotted spinor.
 
%%% Table Spaces, matrices, groups %%%

\begin{table}[t]
\begin{center}
\begin{tabular}{||c||c|c|c|c|c||}
\hhline{|t:=:t:=:=:=:=:=:t|}
\textbf{$G$} & \textbf{$G_D$} & \textbf{Size of $\Delta_{\dot{\a}}$} & \textbf{Size of $v$} & \textbf{$\V$} & \textbf{$\W$} \\
\hhline{|:=::=:=:=:=:=:|}	
$U(N)$ & $U(k)$ & $k \times N + 2k$ & $N \times N + 2k$ & $\Compl^k$ & $\Compl^N$ \\
\hhline{||-||-|-|-|-|-||}
$O(N)$ & $Sp(k)$ & $2k \times N + 4k$ & $N \times N + 4k$ & $\Compl^{2k}$ & $\Real^N$ \\
\hhline{||-||-|-|-|-|-||}
$Sp(N)$ & $O(k)$ & $k \times 2N + 2k$ & $2N \times 2N + 2k$ & $\Real^k$ & $\Compl^{2N}$ \\
\hhline{|b:=:b:=:=:=:=:=:b|}
\end{tabular}
\end{center}
\caption{Spaces, matrices, groups}\label{ADHMTable}
\end{table}

%%% end of table %%%

Let us look closely to the equations \Ref{vCond}. The second equation can be solved for $Q_\a$:
$$
Q_\a(x) = - {(X + x)}^{-2}{(X + x)}_{\a\dot{\a}} S^{\dag\dot{\a}}T(x).
$$
The first equation gives the following condition for $T(x)$:
\begin{equation}
\label{TCond}
T(x)^\dag \left(\Id_\W + S_{\dot{\a}}{(X + x)}^{-2}S^{\dag\dot{\a}}\right)T(x) = \Id_\W.
\end{equation}

The matrix in the brackets is positively defined and therefore there exists a matrix $M(x)$ such that 
\begin{equation}
\label{MCond}
M(x)^\dag M(x) = \Id_\W + S_{\dot{\a}}{(X + x)}^{-2}S^{\dag\dot{\a}}.
\end{equation}
It follows that $g(x) = M(x)T(x) \in U(N)$. Otherwise here we have found the explicit dependence on the gauge group.

\begin{remark}
When we consider group $SO(N)$ or $Sp(N)$ the equations \Ref{vCond}, \Ref{TCond} and \Ref{MCond} are still valid (modulo some minor changes) provided the following convention is accepted:
\begin{itemize}
\item for $SO(N)$ we replace ${(\cdot)}^\dag \mapsto {(\cdot)}^\tr$,
\item for $Sp(N)$ we replace ${(\cdot)}^\dag \mapsto {(\cdot)}^\dag \bJ_N$.
\end{itemize} 
In particular the equation \Ref{MCond} implies $g(x) = M(x)T(x) \in G$.
\end{remark}

%%% Universal bundle %%%

\subsection{Universal bundle}
\label{UniversalBundle}

It is well known \cite{Moore} that a manifold equipped by an almost complex structure and a hermitian metric allows to define a $\mbox{Spin}^\Compl$-structure. Moreover in that case the complexified tangent bundle can be view as $T\M \otimes \Compl \simeq \Hom(S_+ \otimes L,S_-\otimes L)$. Here $S_\pm$ is the spinor bundles of positive and negative chiralities, and $L$ is the square root of the determinant bundle. Even if $S_\pm$ and $L$ do not exist separately their tensor product $S_\pm \otimes L$ does. That is why $S_\pm$ and $L$ is called sometimes virtual bundles.

On the sections of $S_+$ and $S_-$, that is, on dotted and undotted spinors, the maximal torus of the Lorentz group $\Tor_L$ acts as follows: 
$$
\begin{aligned}
\chi^{\dot{\a}} \mapsto {\chi^{\dot{\a}}}' &= {U_+}^{\dot{\a}}{}_{\dot{\b}} \chi^{\dot{\b}} &
&\mbox{and} &
\psi_\a \mapsto \psi_\a' &= {U_-}_\a{}^\b \psi_\b
\end{aligned}
$$
where
$$
\begin{aligned}
U_+ &= \left(\begin{array}{cc}
\e^{-i\eps_+} & 0 \\
0 & \e^{i\eps_+}
\end{array}\right), & 
U_- &= \left(\begin{array}{cc}
\e^{i\eps_-} & 0 \\
0 & \e^{-i\eps_-}
\end{array}\right), &
&\mbox{where} & \eps_\pm &= \frac{\eps_1 \pm \eps_2}{2}.
\end{aligned} 
$$
The complex coordinates transform as:
$$
\begin{aligned}
z_1\mapsto z_1' &= z_1 \e^{i\eps_1} & &\mbox{and} & z_2\mapsto z_2' &= z_2 \e^{i\eps_2},
\end{aligned}
$$
and the sections of $L$ as $s \mapsto s \e^{i\eps_+}$.

Taking into account these properties the ADHM construction can be represented by a following complex:
\begin{equation}
\label{ADHMcomplex}
\V\otimes \L^{-1} \stackrel{\tau}{\longrightarrow} \V \otimes \S_- \oplus \W \stackrel{\sigma}{\longrightarrow} \V\otimes \L
\end{equation}
where 
$$
\begin{aligned}
\tau &= \left(
\begin{array}{c}
B_1 \\
B_2 \\
I
\end{array}
\right), & \sigma &= (B_2,-B_1,J),
\end{aligned}
$$
where $\L$ and $\S_-$ can be viewed as fibers of $L$ and $S_-$ respectively. The ADHM equations \Ref{ADHM} assure that this is indeed a complex.

Now we recall construction of the universal bundle. Let $\Mod$ be the instanton moduli space, given by ADHM construction. Let us introduce local coordinates on $\Mod$: $\{m^I\}$, $I=1,\dots,\dim \Mod$. The tangent space to a point $m \in \Mod$ is spanned by solutions of the linearized self-dual equation \Ref{SDE}. Let us fix a basis of such a solutions: $\{ a^I_\mu(x,m) \}$. Consider now a family of instanton gauge fields parametrized by points of $\Mod$: $A_\mu(x,m)$. We can write
$$
\frac{\pd A_\mu}{\pd m^I} = h_{IJ} a^J_\mu + \Dirac_\mu \a_I
$$ 
where $\a_I$ is a compensating gauge transformation. We can combine it with the connection $A_\mu$ into a one form on $\Real^4\times \Mod $: $\bA = A_\mu \dd x^\mu + \a_I \dd m^I$ which can be seen as a connection of the vector bundle $\E$ over $\Real^4 \times \Mod $ with the fiber $\W$. This bundle is called the universal bundle.

Let $q$ be generic element of the torus $\Tor = \Tor_G \times \Tor_D \times \Tor_L \times \Tor_F$ where the last term $\Tor_F$ is the maximal torus of the flavor group which acts on the solutions of the Dirac equation in the fundamental representation.   

The equivariant Chern character of $\E$ depending on $q$ can be computed as an alternating sum of traces over the cohomologies of the complex \Ref{ADHMcomplex} (see \cite{SWfromInst,SmallInst} for some details). Then we come to the formula
\begin{equation}
\label{ChernE}
\begin{aligned}
\Ch_q(\E) &\equiv \Tr_\E(q) = \Tr_\W(q) + \Tr_\V(q)\Big( \Tr_{\S_-}(q) - \Tr_{\L}(q) - \Tr_{\L^{-1}}(q)\Big) \\
&= \Tr_\W(q) - (\e^{i\eps_1} - 1)(\e^{i\eps_2} - 1)\e^{-i\eps_+} \Tr_\V(q)
\end{aligned}
\end{equation}
where $\Ch_q(\E)$ is the equivariant Chern character. 

The equivariant analogue of the Atiyah-Singer theorem allows us to compute the equivariant index of the Dirac operator. It has the following form
$$
\Ind_q = \sum_{\a} \ep_\a \e^{w_\a} = \int_{\Compl^2} \Ch_q(\E) \Td_q(\Compl^2),
$$
where the sum is taken over all fixed points of the $\Tor$ action and all $\Tor$ action invariant subspaces of the tangent space to a fixed point, $w_\a$ being a weight of this action. In this formula $\Td_q(\Compl^2)$ is the equivariant analogue of the Todd class\footnote{the fact that we should use the Todd class, and not the $\hat{A}$-polynomial, as one could think, follows from the close relation between solutions of the Dirac equation and Dolbeaut cohomology, discussed at the beginning of the section \ref{DiracDbar}.}, which for $\Compl^2 \simeq \Real^4$ has the simple form:
$$
\Td_q(\Compl^2) = \frac{\eps_1\eps_2}{(\e^{i\eps_1} - 1)(\e^{i\eps_2} - 1)}.
$$

The integration can be performed with the help of the DH formula \Ref{DH}. The Hamiltonian of $\Tor_L$ action is $i\eps_1|z_1|^2 + i\eps_2|z_2|^2$. The only fixed point of this action on $\Compl^2$ is the origin. The weights are $i\eps_1$ and $i\eps_2$. Consequently we arrive at
\begin{equation}
\label{sumFund}
\Ind^{\rm{fund}}_q = \sum_{\a} \epsilon_\a \e^{w_\a} = \frac{{\left.\Ch_q(\E)\right|}_{z_1 = z_2 = 0}}{(\e^{i\eps_1} - 1)(\e^{i\eps_2} - 1)}.
\end{equation}

Let us denote the elements of $\Tor_G$, $\Tor_D$ and $\Tor_F$ as follows:
$$
\begin{aligned}
\Tor_G \ni q_G &= \diag \{ ia_1,\dots,ia_N \} \\
\Tor_D \ni q_D &= \diag \{ i\f_1,\dots,i\f_k \} \\
\Tor_F \ni q_F &= \diag \{ i m_1,\dots,i m_{N_f} \}
\end{aligned}
$$
where $a_1,\dots a_N, \f_1,\dots,\f_k, m_1,\dots,m_{N_f}$ are real, $N_f$ being the number of flavors. Then combining \Ref{ChernE} and \Ref{sumFund} we get for $N_f = 1, m = 0$
\begin{equation}
\label{IndSUFund}
\Ind^{\rm{fund}}_q = \sum_{\a} \ep_\a \e^{w_\a} = \frac{1}{(\e^{i\eps_1} - 1)(\e^{i\eps_2} - 1)}\sum_{l=1}^N \e^{ia_l} - \sum_{i=1}^k \e^{i\f_i - i\eps_+}.
\end{equation}

The generalization to $N_f>1$ is straightforward and we obtain:
$$
\Ind^{{\rm{fund}},N_f}_q = \frac{1}{(\e^{i\eps_1} - 1)(\e^{i\eps_2} - 1)}\sum_{f=1}^{N_f}\sum_{l=1}^N \e^{ia_l + im_f} - \sum_{f=1}^{N_f}\sum_{i=1}^k \e^{i\f_i - i\eps_+ + im_f}.
$$

%%% Alternative derivation for the Ch(E) %%%

\subsection{Alternative derivation for $\mathop{\mathrm{Ch}}\nolimits_q(\mathcal{E})$}
\label{DiracDbar}

The derivation of \Ref{IndSUFund} presented in previous section, yet quite general, may seem to be too abstract. Here we present an alternative way to get it. In particular this method allows us to see the origin of all terms which appears in the formula.

Before starting let us recall the relation between Dirac operator on complex manifolds and $\bar\pd$ operator. Define
$$
\bar\pd = \dd\bar{z}_{\bar{1}} \Dirac_{\bar{1}} + \dd\bar{z}_{\bar{2}} \Dirac_{\bar{2}}.
$$
Thanks to ADHM equations \Ref{ADHM} this operator is nilpotent $\bar{\pd}^2 = 0$. Hence the solutions of the Weyl equation \Ref{WEQ} can be naturally associated with Dalbaut cohomologies. The only thing that should be taken into account is the twist by the square root of the determinant bundle. 

Let us turn to the formula \Ref{SolWeyl}. It gives all the solutions of the Weyl equation. In order to specify one solution we need a vector $\xi \in \tilde{\V}$. We wish to have a fermionic solution, since the matter field are supposed to be fermionic. It follows that $\xi$ should be a Grassman number. To show it we put tilde over $\V$. The solution associated to this vector is given by 
$$
\psi^\a_\xi = v^\dag \B^\a \R \xi.
$$
Keeping in mind that for given instanton background (specified by  matrix $\A_{\dot{\a}}$) the matrix  $v$ is defined  by an element of $\G = \{g:\Sphere^4 \to G \}$, we see that a solution of the Weyl equation are labeled by $\G \oplus \tilde{\V}
$.

We stress that it is not the moduli space, since we {\em do not} factor out the group of local gauge transformations $\G$.  

Now we have enough information to reconstruct the equivariant index of the Dirac operator. It is given by the sum of $\Tor$ action weights to fixed points.

Since the gauge transformation $g$ should be $\bar{\pd}$-closed, that is, holomorphic, we conclude that
$$g = \sum_{n_1,n_2\geq 0} g_{n_1,n_2} z_1^{n_1}z_2^{n_2}.$$
The weights of the $\Tor$ action on $\G$ are 
$$
\begin{aligned}
w_\a &= ia_l + in_1\eps_1 + in_2\eps_2, &  l& =1,\dots,N, & n_1,n_2 &\geq 0.
\end{aligned}
$$
The $\Tor$ action on $\tilde{\V}$ is given by $\xi \mapsto \e^{-i\eps_+}q_D\xi$ where $q_D \in \Tor_D$. It follows that the weights are given by
$$
\begin{aligned}
w_\a &= i \f_i - i\eps_+, & i &= 1,\dots,k.
\end{aligned}
$$

Now we recall that the contribution of the fermionic variables comes with $\ep_\a = -1$ (see remark below \Ref{DH}). It implies that the equivariant index equals to
$$
\Ind_q^{\mr{fund}}  = \sum_{l=1}^N \sum_{n_1,n_2 \geq 0} \e^{ia_l + in_1\eps_1 + in_2\eps_2} - \sum_{i=1}^k \e^{i\f_i - i\eps_+}
$$
which is equivalent to \Ref{ChernE} after applying \Ref{sumFund}.

%%% Equivariant index for other groups %%%

\subsection{Equivariant index for other groups}

In a similar way we can find the equivariant index for the fundamental representation of $SO(N)$ and $Sp(N)$.

\microsection{$SO(N)$.} Let $N = 2n + \c$ where $n = [N/2]$ and $\c \equiv N \pmod 2$. Then 
\begin{equation}
\label{IndSOFund}
\Ind_q^{\mr{fund}} = \frac{1}{(\e^{i\eps_1}- 1)(\e^{i\eps_2} - 1)} \left( \c + \sum_{l=1}^n \left(\e^{ia_l} + \e^{-ia_l}\right)\right) - \sum_{i=1}^k \left( \e^{i\f_i -i\eps_+} + \e^{-i\f_i - i\eps_+} \right).
\end{equation}

\microsection{$Sp(N)$.} Let $k = 2n + \c$ where $n = [k/2]$ and $\c \equiv k \pmod 2$. Then
\begin{equation}
\label{IndSpFund}
\Ind_q^{\mr{fund}} = \frac{1}{(\e^{i\eps_1}- 1)(\e^{i\eps_2} - 1)} \sum_{l=1}^N \left( \e^{ia_l} + \e^{-ia_l} \right) - \sum_{i=1}^n \left( \e^{i\f_i-i\eps_+} + \e^{-i\f_i-i\eps_+} + \c \e^{-i\eps_+}\right).
\end{equation}

%%% Other representations and groups %%%

\subsection{Equivariant index for other representations} 

Having computed the equivariant index for the fundamental representation let us turn to others representations. 

\microsection{$SU(N)$.} As it was explained in \cite{SmallInst} the equivariant index for the adjoint representation of $SU(N)$ can be obtained as follows:
\begin{equation}
\label{IndAdjAtiyah}
\begin{aligned}
\Ind^{\rm{adj}}_q &= \sum_{\a}\ep_\a \e^{w_\a} = \int_{\Compl^2} \Ch_q(\E\otimes\E^\ast) \Td_q(\Compl^2) \\
&= \int_{\Compl^2}\Ch_q(\E)\Ch_q(\E^\ast)\Td_q(\Compl^2) = \frac{{\left.\Ch_q(\E)\Ch_q(\E^\ast)\right|}_{z_1=z_2=0}}{(\e^{i\eps_1}-1)(\e^{i\eps_2}-1)}.
\end{aligned}
\end{equation}
We can use the expression \Ref{ChernE} to compute this index. The result is
\begin{equation}
\label{IndSUAdj}
\begin{aligned}
\Ind^{\mr{adj}}_q &=  \frac{1}{(\e^{i\eps_1}-1)(\e^{i\eps_2} - 1)}\left[ N + \sum_{l\neq m}^N \e^{ia_l-ia_m} \right] \\
&- \sum_{i=1}^k\sum_{l=1}^N \left( \e^{i\f_i  - i\eps_+ - ia_l} + \e^{- i\f_i + ia_l - i\eps_+}\right) + k (1-\e^{-i\eps_1})(1-\e^{-i\eps_2})\\
&+ \sum_{i\neq j}^k \left( \e^{i\f_i - i\f_j} + \e^{i\f_i - i\f_j - i\eps_1 - i\eps_2} - \e^{i\f_i - i\f_j - i\eps_1} -  \e^{i\f_i - i\f_j - i\eps_2}\right).
\end{aligned}
\end{equation}

At the same way the indices for symmetric and antisymmetric representations can be obtained. Denote
$$
\begin{aligned}
\Ch_q^{\mr{sym}}(\E) &= \Ch_q(\Sym^2\E), \\
\Ch_q^{\mr{ant}}(\E) &= \Ch_q(\Ant^2\E).
\end{aligned}
$$
If $\Ch_q^{\mr{fund}}(\E) = \sum_\a \ep_\a \e^{w_\a}$ then
\begin{equation}
\label{IndSymAnt}
\Ch_q^{\mr{sym},\mr{ant}}(\E) = \frac{1}{2}\left[ {\left(\Ch_q^{\mr{fund}}\right)}^2 \pm \Ch^{\mr{fund}}_{q^2}\right] = \frac{1}{2}\left[ {\Big( \sum_\a \ep_\a \e^{w_a} \Big)}^2 \pm \sum_\a \ep_\a \e^{2w_\a}\right].
\end{equation}
We can now apply the analogue of \Ref{IndAdjAtiyah} to compute the equivariant index for these representations. The result is the following:
\begin{equation}
\label{IndSUSym}
\begin{aligned}
\Ind_q^{\mr{sym}} &= \frac{1}{(\e^{i\eps_1} - 1)(\e^{i\eps_2} - 1)} \sum_{l\leq m \leq N} \e^{ia_l + ia_m} \\
&- \sum_{l=1}^N \sum_{i=1}^k \e^{ia_l + i\f_i - i\eps_+} - \sum_{i=1}^k \left( \e^{2i\f_i - i\eps_1} + \e^{2i\f_i - i\eps_2}\right) \\ 
&+ \sum_{i<j\leq k} \left( \e^{i\f_i + i\f_j} + \e^{i\f_i + i\f_j - i\eps_1 - i\eps_2} - \e^{i\f_i + i\f_j - i\eps_1} -  \e^{i\f_i + i\f_j - i\eps_2}\right),
\end{aligned}
\end{equation}
\begin{equation}
\label{IndSUAnt}
\begin{aligned}
\Ind_q^{\mr{ant}} &= \frac{1}{(\e^{i\eps_1} - 1)(\e^{i\eps_2} - 1)} \sum_{l <  m \leq N} \e^{ia_l + ia_m} \\
&- \sum_{l=1}^N \sum_{i=1}^k \e^{ia_l + i\f_i - i\eps_+} + \sum_{i=1}^k \left( \e^{2i\f_i} + \e^{2i\f_i - i\eps_1 - i\eps_2}\right) \\ 
&+ \sum_{i<j\leq k} \left( \e^{i\f_i + i\f_j} + \e^{i\f_i + i\f_j - i\eps_1 - i\eps_2} - \e^{i\f_i + i\f_j - i\eps_1} -  \e^{i\f_i + i\f_j - i\eps_2}\right).
\end{aligned}
\end{equation}

%%% Table Group theoretical data %%%

\begin{table}[t]
\begin{center}
\begin{tabular}{||c||c|c|c|c|c|c|c||}
\hhline{|t:=:t:=:=:=:=:=:=:=:t|}
\textbf{Group} & \textbf{$h$} & \textbf{$h^\vee$} & \textbf{$|W|$} & \textbf{Adjoint}  & \textbf{$\ell_{\mr{adj}}$} & \textbf{$\ell_{\mr{sym}}$} & \textbf{$\ell_{\mr{ant}}$}\\
\hhline{|:=::=:=:=:=:=:=:=:|}	
$A_n:SU(n+1)$  & $n+1$ &$n+1$ & $(n+1)!$ & $\mbox{fund}\otimes\mbox{fund}^\ast$ & $2n+2$ & $2n+4$ & $2n$\\
\hhline{||-||-|-|-|-|-|-|-||}
$B_n:SO(2n+1)$  & $2n$ & $2n-1$ & $2^n n!$ & $\Ant^2\mbox{fund}$ & $2n-1$  & $2n + 3$& $2n - 1$\\
\hhline{||-||-|-|-|-|-|-|-||}
$C_n:Sp(n)$ & $2n$ & $n+1$ & $2^n n!$ & $\Sym^2\mbox{fund}$ & $2n+2$ & $2n+2$ & $2n-2$\\
\hhline{||-||-|-|-|-|-|-|-||}
$D_n:SO(2n)$ & $2n-2$ & $2n-2$ & $2^{n-1}n!$ & $\Ant^2\mbox{fund}$ & $2n-2$  & $2n + 2$& $2n - 2$\\
\hhline{|b:=:b:=:=:=:=:=:=:=:b|}
\end{tabular}
\end{center}
\caption{Group theoretical data \cite{BourbakiLie}}\label{GroupTh}
\end{table}

%%% end of table %%%

\begin{remark}
Using formulae \Ref{IndSOFund} and \Ref{IndSpFund} we can write similar expression for orthogonal and symplectic gauge group models.
\end{remark}

%%% Partition function %%%

\subsection{Partition function}

Now we are ready to write the expression for the partition function \Ref{Z}. First we note that since the tangent space to a point belonging to a bundle is a direct sum of the tanget space to the point of the base and the tangent space to the point of the fiber. Taking into account the statistics of the fields (recall that the Yang-Mills connection $A_\mu$ belongs to the adjoint representation of the gauge group) we can write
\begin{equation}
\label{AddInd}
\Ind_q = \Ind_q^{\mr{adj,gauge}} - \sum_{r \in \mr{reps}} \Ind_q^{r,\mr{matter}}.
\end{equation}
The transformation \Ref{TrWeights} converts the sum to a product. The last step consists of taking the dual group invariant part of the expression. This task can be accomplished by taking the integral over the dual group. This integral can be reduced to the integral over the torus $\Tor_D$. The price we pay is the Weyl-Vandermond factor in the measure and the order of the  Weyl group of the dual group which we divide the integral on. See \cite{DeformationInstanton,GeneralizedInst,ABCD} for more details.

Let us realize this program step-by-step. Compute first the DH products \Ref{TrWeights} for (almost) all cases allowed by the asymptotic freedom. We will consider all the matter representations contained in a tensor power of the fundamental representation. For $SU(N)$ we can get all the representations in such a way. However for other group this is not the case. For example for $SO(N)$ we will miss some spinor representations. We should find the solutions of the equation $\beta \geq 0$ where $\beta$ is defined by the righthand side of \Ref{ass}. Using Table \ref{GroupTh} we get the following list (Table \ref{Models}).

%%% Table Models %%%

\begin{table}
\begin{itemize}
\item \underline{$SU(N)$}:
\begin{itemize}
\item $N_f$ fundamental multiplets, $N_f \leq 2N$,
\item 1 antisymmetric multiplet and $N_f$ fundamental, $N_f \leq N + 2$,
\item 1 symmetric multiplet and $N_f$ fundamental, $N_f \leq N-2$,
\item 2 antisymmetric and $N_f$ fundamental, $N_f \leq 4$,
\item 1 symmetric and 1 antisymmetric multiplet,
\item 1 adjoint multiplet.
\end{itemize}
\item \underline{$SO(N)$}:
\begin{itemize}
\item $N_f$ fundamental multiplet, $N_f \leq N-2$,
\item 1 adjoint multiplet.
\end{itemize}
\item \underline{$Sp(N)$}:
\begin{itemize}
\item $N_f$ fundamental multiplet, $N_f \leq N+2$,
\item 1 antisymmetric multiplet and $N_f$ fundamental, $N_f \leq 4$,
\item 1 adjoint multiplet.
\end{itemize}
\end{itemize}
\caption{Models allowed by the asymptotic freedom}\label{Models}
\end{table}

%%% end of table %%%

Here we give the expression for the building blocks which are necessary to construct all the cases listed above. In all formulae we set $\eps = 2\eps_+ = \eps_1 + \eps_2$.

%%% SU(N) case %%%

\microsection{$SU(N)$ case.} Denote
$$
\begin{aligned}
\Delta_\pm(x) &= \prod_{i<j\leq k} \Big( (\f_i \pm \f_j)^2 - x^2\Big) \\
\P(x) &= \prod_{l=1}^N (x-a_l).
\end{aligned}
$$
Then
\begin{equation}
\label{zSUFund}
z_k^{\mr{fund}}(q) = Z^{\mr{fund}}_{\mr{pert}}(q) \times  \prod_{i=1}^k (\phi_i+m  -\eps_+), 
\end{equation}
\begin{equation}
\label{zSUAdjGauge}
z_k^{\mr{adj,gauge}}(q) = Z^{\mr{adj,gauge}}_{\mr{pert}}(q) \times \frac{\eps^k}{\eps_1^k\eps_2^k}\frac{\Delta_-(0)\Delta_-(\eps)}{\Delta_-(\eps_1)\Delta_-(\eps_2)}\prod_{i=1}^k \frac{1}{\P(\f_i+\eps_+)\P(\f_i-\eps_+)}, 
\end{equation}
\begin{equation}
\label{zSUAdjMatter}
\begin{aligned}
z_k^{\mr{adj,matter}}(q) &= Z^{\mr{adj,matter}}_{\mr{pert}}(q) \times  \frac{(m-\eps_1)^k(m-\eps_2)^k}{(m-\eps)^k m^k} \frac{\Delta_-(m-\eps_1)\Delta_-(m-\eps_2)}{\Delta_-(m)\Delta_-(m-\eps)} \\
&\times \prod_{k=1}^k \P(\f_i - m + \eps_+)\P(\f_i + m -\eps_+), 
\end{aligned}
\end{equation}
\begin{equation}
\label{zSUSym}
\begin{aligned}
z_k^{\mr{sym}}(q) &= Z^{\mr{sym}}_{\mr{pert}}(q) \times \frac{\Delta_+(m-\eps_1)\Delta_+(m-\eps_2)}{\Delta_+(m)\Delta_+(m-\eps)} \\
&\times \prod_{i=1}^k ( 2\f_i +m -\eps_1)(2\f_i + m -\eps_2)\P(-\f_i -m + \eps_+), 
\end{aligned}
\end{equation}
\begin{equation}
\label{zSUAnt}
z_k^{\mr{ant}}(q) = Z^{\mr{ant}}_{\mr{pert}}(q) \times \frac{\Delta_+(m-\eps_1)\Delta_+(m-\eps_2)}{\Delta_+(m)\Delta_+(m-\eps)}\prod_{i=1}^k\frac{\P(-\f_i - m + \eps_+)}{(2\f_i + m)(2\f_i +m - \eps)}.
\end{equation}
\begin{remark}
The term $Z_{\mr{pert}}(q)$ comes from the first terms in \Ref{IndSUFund}, \Ref{IndSUAdj}, \Ref{IndSUSym}, \Ref{IndSUAnt} respectively. Under the transformation \Ref{TrWeights} these terms become the infinite products to be regularized. It can be shown \cite{SWfromInst,SWandRP} that after the proper time regularization they give precisely the perturbative contribution to the prepotential \Ref{PrepPert} (in the $\eps_1,\eps_2\to 0$ limit, see section \ref{SU(N)PureYM}). In that follows we will drop this term in all calculations and restore it, if ever, only in the final result. 
\end{remark}

To find similar expressions for $SO(N)$ and $Sp(N)$ we use \Ref{IndSOFund}, \Ref{IndSpFund}, and \Ref{IndSymAnt}. The result is the following.

%%% SO(N) case %%%

\microsection{$SO(N)$ case.} Denote
$$
\begin{aligned}
\Delta(x) &= \prod_{i<j\leq k} \Big( (\f_i + \f_j)^2 - x^2 \Big) \Big( (\f_i -\f_j)^2 - x^2\Big), \\
\P(x) &= x^\c \prod_{l=1}^n (x^2 - a_l^2).
\end{aligned}
$$
Then
\begin{equation}
\label{zSOFund}
z_k^{\mr{fund}}(q) = \prod_{i=1}^k ((m-\eps_+)^2 - \f_i^2),
\end{equation}
\begin{equation}
\label{zSOAdjGauge}
z_k^{\mr{adj,gauge}}(q) = \frac{\eps^k}{\eps^k_1\eps^k_2} \frac{\Delta(0)\Delta(\eps)}{\Delta(\eps_1)\Delta(\eps_2)}\prod_{i=1}^k \frac{4\phi_i^2 (4\f_i^2 - \eps^2)}{\P(\f_i + \eps_+)\P(\f_i - \eps_+)}, 
\end{equation}
\begin{equation}
\label{zSOAdjMatter}
\begin{aligned}
z_k^{\mr{adj,matter}}(q) &= \frac{(m-\eps_1)^k(m-\eps_2)^k}{m^k(m-\eps)^k}\frac{\Delta(m-\eps_1)\Delta(m-\eps_2)}{\Delta(m)\Delta(m-\eps)} \\
&\times \prod_{i=1}^k \frac{\P(\f_i + m - \eps_+)\P(\f_i -m + \eps_+)}{(4\f_i^2 - m^2)(4\f_i^2 - (m-\eps)^2)}.
\end{aligned}
\end{equation}

%%% Sp(N) case %%%

\microsection{$Sp(N)$ case.} Denote
$$
\begin{aligned}
\Delta(x) &= \prod_{i<j\leq n} \Big( (\f_i + \f_j)^2 - x^2\Big)\Big( (\f_i - \f_j)^2 - x^2\Big), \\
\P(x) &= \prod_{l=1}^N (x^2 - a_l^2).
\end{aligned}
$$
Then
\begin{equation}
\label{zSpFund}
z_k^{\mr{fund}}(q) = (m-\eps_+)^\chi\prod_{i=1}^n ( (m-\eps_+)^2 - \phi_i^2),
\end{equation}
\begin{equation}
\label{zSpAdjGauge}
\begin{aligned}
z_k^{\mr{adj,gauge}}(q) &= \frac{\eps^n}{\eps_1^n\eps_2^n} {\left[ \frac{1}{\eps_1\eps_2 \prod_{l=1}^N (\eps_+^2 - a_l^2)} \prod_{i=1}^n \frac{\f_i^2(\f_i^2 - \eps^2)}{(\f_i^2 - \eps_1^2)(\f_i^2 -\eps_2^2)}\right]}^\c \\
&\times \frac{\Delta(0)\Delta(\eps)}{\Delta(\eps_1)\Delta(\eps_2)}\prod_{i=1}^n\frac{1}{\P(\f_i - \eps_+)\P(\f_i + \eps_+)(4\f_i^2 - \eps_1^2)(4\f_i^2 - \eps_2^2)}, 
\end{aligned}
\end{equation}
\begin{equation}
\label{zSpAdjMatter}
\begin{aligned}
z_k^{\mr{adj,matter}}(q) &= \frac{(m-\eps_1)^n(m-\eps_2)^n}{m^n(m-\eps)^n}\frac{\Delta(m-\eps_1)\Delta(m-\eps_2)}{\Delta(m)\Delta(m-\eps)}\\
&{\times\left[ (m-\eps_1)(m-\eps_2) \prod_{l=1}^N\Big((m-\eps_+)^2 - a_l^2\Big)\prod_{i=1}^n\frac{(\f_i^2 - (m-\eps_1)^2)(\f_i^2 - (m-\eps_2)^2)}{(\f_i^2 - m^2)(\f_i^2 - (m-\eps)^2)}\right]}^\c \\
&\times \prod_{i=1}^n\P(\f_i +m - \eps_+)\P(\f_i -m + \eps_+)\prod_{s=1}^2(4\f_i^2 - (m - \eps_s)^2), 
\end{aligned}
\end{equation}
\begin{equation}
\label{zSpAnt}
\begin{aligned}
z_k^{\mr{ant}}(q) &= \frac{(m-\eps_1)^n(m-\eps_2)^n}{m^n(m-\eps)^n}\frac{\Delta(m-\eps_1)\Delta(m-\eps_2)}{\Delta(m)\Delta(m-\eps)}  \\
&{\times \left[ \frac{ \prod_{l=1}^N((m-\eps_+)^2 - a_l^2)}{m(m-\eps)}\prod_{i=1}^n\frac{(\f_i^2 - {(m-\eps_1)}^2)(\f_i^2 - {(m-\eps_2)}^2)}{(\f_i^2 - m^2)(\f_i^2 - {(m-\eps)}^2)}\right]}^\c \\
&\times\prod_{i=1}^n \frac{\P(\f_i + m -\eps_+)\P(\f_i - m + \eps_+)}{(4\f_i^2 - m^2) (4\f_i^2 - {(m-\eps)}^2)}.
\end{aligned} 
\end{equation}
Now we should perform the integration over $\Tor_D$. The order of the Weyl group of the dual group $|W_D|$ can be found in Table \ref{GroupTh}. We arrive to the following expression:
\begin{equation}
\label{partition}
Z_k(a,m,\eps) = \frac{1}{|W_D|} \oint \prod_{i=1}^k \frac{\dd \f_i}{2\pi i} z_k^{\mr{adj,gauge}} (a,\eps,\f)\prod_{r \in \mr{reps}} z_k^{r,\mr{matter}}(a,m_r,\eps,\f)
\end{equation}

\begin{remark}
The expressions for the adjoint representation integrand $z_k(q)$ for $SO(N)$ and $Sp(N)$ coincides with the expressions which can be obtained from the direct analysis of the instanton moduli space for these groups \cite{ABCD}.
\end{remark}

To compute the contour integral we need a contour bypassing prescription. It can be obtained, as explained in \cite{ABCD}, by considering the four dimensional theory as a limit of a five dimensional theory, where the complexified torus $\Tor_\Compl$ acts on. As a result we obtain $(\eps_1,\eps_2,m) \mapsto (\eps_1 + i0,\eps_2 + i0, m - i0)$ prescription. It worth noting that the prescription for masses $m$ coincides with the Feynman prescription for bypassing the physical poles. The contour can be closed on the upper or lower complex halfplain. The choice is irrelevant since the residue at infinity of the integrand  vanishes. 
%%% 1-instanton corrections %%%

\subsection{1-instanton corrections and residue functions}
\label{1instComp}

Formula \Ref{partition}, yet far from the final result, allows, however, to perform various checks. In particular, we can compare this formula against the known one instanton corrections.

After the work of Seiberg and Witten \cite{SeibergWitten} the 1-instanton corrections was computed for numerous combinations of (classical) groups and matter content. In particular, in references \cite{NaculichSUAntFund,EnnesSUSymFund,EnnesSUSymAnt,EnnesSU2AntFund,HKP-SU,SOandSp,CMandSW,EllipticMod} it was done for all cases allowed by asymptotic freedom.

In \cite{EnnesSU2AntFund,EnnesMasterFunc,EllipticMod,MtheoryTested} it was pointed out that in all cases the one instanton corrections can be described with the help of a rational function $S(x)$ referred as a {\em master function} or {\em residue function}. This function appears in the hyperelliptic truncation of the Seiberg-Witten curve as follows:
\begin{equation}
\label{HyperEllipticTrunc}
y(z) + \frac{1}{y(z)} = \frac{1}{\sqrt{S(z)\Lambda^\b}}.
\end{equation}

The rules to construct such a function was proposed in \cite{EllipticMod,MtheoryTested}. We have put them to the Table \ref{Srules}.

%%% Begin Table S-rules %%%

\begin{table}
\begin{center}
\begin{tabular}{||c|c||c||}
\hhline{|t:=:=:t:=:t|}
\textbf{Group} & \textbf{Multiplet} & \textbf{Factor of $S(x)$}\\
\hhline{|:=:=::=:|}	 
& Adjoint, gauge & $\dfrac{1}{\prod_{l=1}^N {(x-a_l)}^2}$\\
\hhline{||~|-||-||}
& Fundamental & $x+m$\\
\hhline{||~|-||-||}
$SU(N)$ & Symmetric & ${(2x + m)}^2 \prod_{l=1}^N(x + a_l + m)$\\
\hhline{||~|-||-||}
& Antisymmetric & $\dfrac{1}{{(2x + m)}^2}\prod_{l=1}^N (x + a_l + m)$\\
\hhline{||~|-||-||}
& Adjoint, matter & $\prod_{l=1}^N ({(x - a_l)}^2 - m^2)$\\
\hhline{|:=:=::=:|}
& Adjoint, gauge & $\dfrac{x^{4-2\c}}{\prod_{l=1}^n {(x^2 - a_l^2)}^2}$\\
\hhline{||~|-||-||}
$SO(2n + \c)$ & Fundamental & $x^2 - m^2 $\\
\hhline{||~|-||-||}
$(\c = 0,1)$& Adjoint, matter & $\dfrac{{(x^2 - m^2)}^\c}{4x^2 - m^2} \prod_{l=1}^n ({(x + m)}^2 - a_l^2)({(x - m)}^2 -a_l^2)$\\
\hhline{|:=:=::=:|}
& Adjoint, gauge & $\dfrac{1}{x^4 \prod_{l=1}^N{(x^2 - a_l^2)}^2}$\\
\hhline{||~|-||-||}
$Sp(N)$ & Fundamental & $x^2 -m^2$\\
\hhline{||~|-||-||}
& Antisymmetric &  $\dfrac{\prod_{l=1}^N({(x+m)}^2 - a_l^2)({(x-m)}^2 - a_l^2)}{{(4x^2 - m^2)}^2}$\\
\hhline{||~|-||-||}
& Adjoint, matter & ${(4x^2 - m^2)}^2 \prod_{l=1}^N({(x+m)}^2 - a_l^2)({(x-m)}^2 - a_l^2)$\\ 
\hhline{|b:=:=:b:=:b|}
\end{tabular}
\end{center}
\caption{$S(x)$ building blocks}\label{Srules}
\end{table}

%%% end of table %%%

The residue function  has double and quadratic poles. Denote the corresponding ``residues'' as follows:
$$
\begin{aligned}
S(x) &\sim \frac{S_2(x)}{(x-x_0)^2}, & S(x) &\sim \frac{S_4(x)}{(x-y_0)^4}.
\end{aligned}
$$

Then in many cases the one instanton corrections are given by 
\begin{equation}
\label{Prep1}
\Prep_1(a,m) = \sum_{l=1}^N S_2(a_l).
\end{equation}
If the model contains  one antisymmetric representation of $SU(N)$ or the adjoint of $SO(N)$ one should add to \Ref{Prep1} term $-2S_2(-m/2)$, where $m$ is the mass of corresponding matter multiplet. For two antisymmetric multiplets of $SU(N)$ with masses $m_1$ and $m_2$ one adds $-2S_2(-m_1/2)-2S_2(-m_2/2)$.

Finally for the group $Sp(N)$ we have a quite different expression. One instanton corrections for all matter multiplets is given by
$$
\Prep_1(a,m) =\sqrt{S_4(0)}.
$$

The aim of this section is to show how the notion of the residue function naturally appears in our approach. This analysis allows us to state that 1-instanton corrections computed by our method match with 1-instanton corrections computed from $M$-theory curves.

Put $k=1$. The 1-instanton contribution to the partition function \Ref{partition} is given by
\begin{equation}
\label{Z1}
Z_1(a,m,\eps) = \oint \frac{\dd\f}{2\pi i} z_1(a,m,\eps,\f).
\end{equation}
The 1-instanton correction to the prepotential can be extracted from $Z_1(a,m,\eps)$ according to
\begin{equation}
\label{F1fromZ1}
Z_1(a,m,\eps) = \frac{1}{\eps_1\eps_2}\Prep_1(a,m)  + \dots,
\end{equation}
where ``$\dots$'' denotes all terms containing nonnegative powers of $\eps_1,\eps_2$. Combining these two formulae we get
$$
\Prep_1(a,m) = \lim_{\eps_1,\eps_2 \to 0} \eps_1\eps_2 \oint \frac{\dd\f}{2\pi i}z_1(a,m,\eps,\f).
$$

%%% Table z1 building blocks %%%

\begin{table}
\begin{center}
%\begin{sideways}
\begin{tabular}{||c|c||c||}
\hhline{|t:=:=:t:=:t|}
\textbf{Group} & \textbf{Multiplet} & \textbf{Factor of $z_1(a,m,\eps,\f)$}\\
\hhline{|:=:=::=:|}	 
& Adjoint, gauge & $\dfrac{\eps}{\eps_1\eps_2} \dfrac{1}{\prod_{l=1}^N ((\f-a_l)^2 - \eps_+^2)}$\\
\hhline{||~|-||-||}
& Fundamental & $\f + m -\eps_+$\\
\hhline{||~|-||-||}
$SU(N)$ & Symmetric & $(2\f + m -\eps_1)(2\f + m -\eps_2)\prod_{l=1}^N(\f + a_l + m - \eps_+)$\\
\hhline{||~|-||-||}
& Antisymmetric & $\dfrac{\prod_{l=1}^N (\f + a_l + m - \eps_+)}{(2\f + m)(2\f + m - \eps)}$\\
\hhline{||~|-||-||}
& Adjoint, matter & $\dfrac{(m-\eps_1)(m-\eps_2)}{(m-\eps)m} \prod_{l=1}^N ((\f - a_l)^2 - (m - \eps_+)^2)$\\
\hhline{|:=:=::=:|}
& Adjoint, gauge & $\dfrac{\eps}{\eps_1\eps_2}\dfrac{4\f^2 (4\f^2 - \eps^2)}{(\f^2 - \eps_+^2)^\c\prod_{l=1}^n ((\f+\eps_+)^2 - a_l^2)((\f-\eps_+)^2 - a_l^2)}$\\
\hhline{||~|-||-||}
$SO(2n + \c)$ & Fundamental & $(m- \eps_+)^2 - \f^2$\\
\hhline{||~|-||-||}
$(\c = 0,1)$& Adjoint, matter & $\dfrac{(m-\eps_1)(m-\eps_2)}{m(m-\eps)}(\f^2 - (m-\eps_+^2)^2)^\c$ \\
& &$\times \dfrac{\prod_{l=1}^n ((\f + m - \eps_+)^2 - a_l^2)((\f - m + \eps_+)^2 -a_l^2)}{(4\f^2 - m^2)(4\f^2 - (m-\eps)^2)}$\\
\hhline{|:=:=::=:|}
& Adjoint, gauge & $\dfrac{1}{\eps_1\eps_2\prod_{l=1}^N(\eps_+^2 - a_l^2)}$\\
\hhline{||~|-||-||}
$Sp(N)$ & Fundamental & $(m-\eps_+)$\\
\hhline{||~|-||-||}
& Antisymmetric & $\dfrac{\prod_{l=1}^N ((m-\eps_+)^2 - a_l^2)}{m(m-\eps)}$\\
\hhline{||~|-||-||}
& Adjoint, matter & $(m-\eps_1)(m-\eps_2)\prod_{l=1}^N ((m-\eps_+)^2 - a_l^2)$\\ 
\hhline{|b:=:=:b:=:b|}
\end{tabular}
%\end{sideways}
\end{center}
\caption{$z_1(a,m,\eps,\f)$ building blocks}\label{z1rules}
\end{table}

%%% end of table %%%

Analysis of \Ref{zSUFund}, \Ref{zSUAdjGauge}, \Ref{zSUAdjMatter}, \Ref{zSUAnt}, \Ref{zSUSym}, \Ref{zSOFund}, \Ref{zSOAdjGauge}, \Ref{zSOAdjMatter}, \Ref{zSpFund}, \Ref{zSpAdjGauge}, \Ref{zSpAdjMatter}, and \Ref{zSpAnt} together with \Ref{partition} shows that one can establish the rule to construct $z_1(a,m,\eps,\f)$ (see Table \ref{z1rules}). 

First observation is that that for $SU(N)$ and $SO(N)$ the following equality holds:
$$
\lim_{\eps_1,\eps_2\to0}\frac{\eps_1\eps_2}{\eps}z_1(a,m,\eps,\f) = S(\f).
$$
Hence one can call $z_1(a,m,\eps,\f)$ a deformed residue function. Using the properties of the contour integration
$$
\begin{aligned}
\oint \frac{\dd \f}{2\pi i} \frac{1}{(\f - x_0 - \eps_+)(\f - x_0 +\eps_+)} = \frac{1}{\eps}
\end{aligned}
$$
we arrive to the rule announced after \Ref{Prep1}.

\begin{remark}
For $Sp(N)$ the integrand does not depend on $\f$. It means that for $Sp(N)$ the one instanton corrections are given by
$$
\Prep_1(a,m) = \lim_{\eps_1,\eps_2\to 0} \eps_1\eps_2z_1(a,m,\eps).
$$
The rule for the residue function proposed in \cite{EllipticMod,MtheoryTested} are such that 
$$\sqrt{S_4(0)} = \lim_{\eps_1,\eps_2\to 0}\eps_1\eps_2 z_1(a,m,\eps).$$
This proves the validity of our formulae in the case of $Sp(N)$.
\end{remark}

The method of residue function, yet simple for $k=1$ case, seems to be difficult to generalize to other ($k>1$) cases. The reason is both the complexity of \Ref{Z1} and \Ref{F1fromZ1} when $k>1$. For example \Ref{Z1} generalizes as follows (for $SU(N)$ and $SO(N)$, the $Sp(N)$ case should be considered separately):
$$
Z_k(a,m,\eps) = \oint \prod_{i=1}^k \frac{\dd \f_i}{2\pi i} \EuScript{R}(\f) \prod_{i=1}^k z_1(a,m,\eps,\f_i)
$$
where $\EuScript{R}(\f)$ is a ratio of $\Delta$'s products. The integral can be computed by hands in low $k$ case. For example, it was done in \cite{WyllMar} for $k \leq 3$ for $SO(N)$ and $Sp(N)$ pure Yang-Mills theories and for $k\leq 2$ for symmetric and antisymmetric representations of $SU(N)$. Also these integrals can be computed for general $k$ in the case of $SU(N)$ (fundamental and adjoint representations, \cite{SWfromInst}). See the discussion in \cite{WyllMar} of what happens in the case of other classical groups.

%%% Episode II: Saddel point equation %%%

\section{Episode II: Saddle point equation}
\label{II}

The formal expression \Ref{partition} allows, in principle, to compute all the instanton correction. However, there are two objection: first, for general group and representation this is not known how to rewrite this integral as a sum over the residues of the deformed residue function $z_1(a,m,\eps,\f)$. Second objection comes from the fact, that the representation of the prepotential by a formal series on $\Lambda$ makes its analytical properties obscure. In particular, it is not clear how the prepotential could be analytically continued beyond the convergence radius. 

Fortunately, the Seiberg-Witten theory \cite{SeibergWitten} can answer to the second question. Our goal in this section is to explain how the Seiberg-Witten data can be extracted from \Ref{partition}.

%%% Thermodynamical (classical) limit %%%

\subsection{Thermodynamical (classical) limit}

In \cite{SWandRP} the general method to extract the Seiberg-Witten data was proposed. The idea is the following. The prepotential can be obtained from the partition function $Z(a,m,\eps,\Lambda)$ in the limit $\eps_1,\eps_2\to 0$ (see \Ref{Z}). One can show that in this limit the main contribution to the partition function comes from $k \sim \dfrac{1}{\eps_1\eps_2}$. It follows that in order to extract Seiberg-Witten data we don't need to examine the whole series \Ref{Z}. It is sufficient to consider the expression \Ref{partition} taken in the limit $k \to\infty$.
 
In this limit the multiple integral on $\f_i$ becomes Feynman integral over the density of $\f_i$'s. Each $\f_i$ can be seen as a physical quantity which corresponds to a ``particle''. The instanton number $k$ plays the role of the number of such a ``particles''. Another point of view is to consider the inverse instanton number as a Plank constant in a quantum mechanical problem. The expression \Ref{partition} becomes the partition function of a system, described by a Hamiltonian, depending of the $\f_i$'s density. 

In the thermodynamical (classical) limit $k\to\infty$ this partition function can be computed by the saddle point approximation. It means that the main contribution is given by a classical configuration (we put aside the question of existence and uniqueness of such a configuration). The prepotential becomes the ``free energy'' in this context. As we shall see the Seiberg-Witten data appears naturally when we solve the equation of motion (saddle point equation).

After this short introduction let us pass to the concrete computations. First we note that the thermodynamical (or quantum mechanical) problem is formulated by means of the action (Hamiltonian). The integrand in the Feynman integral generically has the form $\e^{-\frac{1}{\eps_1\eps_2}H}$. Therefore we should convert the integrand of \Ref{partition} into the similar form. Keeping in mind the origin of this integrand (formula \Ref{TrWeights}) we can obtain a mnemonic rule to compute the Hamiltonian $H$ directly from the equivariant index of the Dirac operator:
$$
\Ind_q = \sum_{\a} \ep_\a \e^{w_\a} \mapsto \prod_{\a} {w_\a}^{\ep_\a}= \exp\left\{\sum_\a \ep_\a \ln w_\a\right\} \mapsto  H_{\eps_1,\eps_2} = - \eps_1\eps_2\sum_{\a} \ep_\a \ln |w_\a|.
$$

However, the Hamiltonian defined above contains much more information we need. Namely it can be represented as a series over the nonnegative powers of $\eps_1$ and $\eps_2$. The only contribution relevant in the thermodynamical limit comes from the terms independent of $\eps_1$ and $\eps_2$. Therefore the expression for the Hamiltonian can be rewritten as follows:
\begin{equation}
\label{HamClass}
\Ind_q = \sum_{\a}\ep_\a \e^{w_\a} \mapsto H = - \lim_{\eps_1,\eps_2\to 0} \eps_1\eps_2 \sum_\a \ep_\a \ln |w_\a|.
\end{equation}

Taking into account the additivity of the equivariant index \Ref{AddInd} we conclude that 
$$
H = H^{\mr{adj,gauge}} + \sum_{r\in \mr{reps}} H^{r,\mr{matter}}.
$$

\begin{remark}
We have just established a rule to represent $Z_k(a,m,\eps)$ given by \Ref{partition} as an exponent of a sum of $\ln|w_\a|$'s. We can ask now what will change if we multiply $Z_k(a,m,\eps)$ by $\Lambda^{k\b}$. The answer is that we should replace $\ln|w_\a|$ with $\ln\left|\dfrac{w_\a}{\Lambda}\right|$.
\end{remark}

%%% SU(N) case, pure Yang-Mills theory %%%

\subsection{$SU(N)$ case, pure Yang-Mills theory}

\label{SU(N)PureYM}

Let us consider in some details the simplest case: the $SU(N)$ theory without matter multiplets. The weights are given by \Ref{IndSUAdj}. 

Let us show how the first term in \Ref{IndSUAdj} gives the perturbative correction to the prepotential \cite{SWandRP}.  

As we have already motioned, the transformation \Ref{TrWeights} can be seen as the proper time regularization. It is given by the formula
$$
\e^{i\langle x,w_\a(p)\rangle} \mapsto {\left. \frac{\dd}{\dd s}\right|}_{s=0}\frac{\Lambda^s}{\Gamma(s)} \int_0^\infty \frac{\dd t}{t} t^s \e^{i\langle tx,w_\a(p)\rangle} = - \ln \left|\frac{\langle x,w_\a(p)\rangle}{\Lambda}\right|.
$$

It follows that the contribution of the first term of \Ref{IndSUAdj} to the Hamiltonian \Ref{HamClass} is given by
$$
\lim_{\eps_1,\eps_2\to 0}\eps_1\eps_2 \sum_{l,m=1}^N \g_{\eps_1,\eps_2}(a_l - a_m,\Lambda)
$$
where 
$$
\g_{\eps_1,\eps_2} (x,\Lambda) = {\left. \frac{\dd}{\dd s}\right|}_{s = 0} \frac{\Lambda^s}{\Gamma(s)}\int_0^\infty \frac{\dd t}{t} t^s \frac{\e^{itx}}{(1-\e^{i\eps_1 t})(1 - \e^{i\eps_2 t})}.
$$
The $\eps$ expansion of $\gamma_{\eps_1\eps_2}$ is given by 
$$
\g_{\eps_1,\eps_2}(x,\Lambda) = \frac{1}{\eps_1\eps_2} \k_\Lambda(x) + \dots,
$$
where ``$\dots$'' are terms finite in the thermodynamical limit and
$$
\k_\Lambda(x) = \frac{1}{2} x^2 \left(\ln \left|\frac{x}{\Lambda}\right| - \frac{3}{2}\right).
$$
For more properties of $\g_{\eps_1,\eps_2}(x,\Lambda)$ see Appendix A in \cite{SWandRP}.

Finally the contribution to the Hamiltonian of the first term is given by
$$
\sum_{l\neq m} \k_\Lambda(a_l - a_m) = \sum_{l\neq m} \frac{1}{2} {(a_l - a_m)}^2 \left(\ln \left|\frac{a_l - a_m}{\Lambda}\right| - \frac{3}{2}\right).
$$

In this expression we can recognize the perturbative part of the prepotential \Ref{PrepPert}. It explains the remark after \Ref{zSUAnt}.

To handle the last line in \Ref{IndSUAdj} we use the following identity:
$$
f(0) + f(\eps_1+\eps_2) - f(\eps_1) - f(\eps_2) = \eps_1\eps_2 f''(0)+ \dots,
$$
where ``$\dots$'' are the higher $\eps$-terms. It gives
%\begin{multline*}
$$
\ln(\f_i - \f_j) + \ln (\f_i - \f_j - \eps) -  \ln (\f_i - \f_j -\eps_2) - \ln (\f_i - \f_j - \eps_1) = -\eps_1\eps_2 \frac{1}{{(\f_i - \f_j)}^2} + \dots.
$$
%\end{multline*}

Finally with the help of $\Ref{rhoSU}$ we have the following expression for the Hamiltonian:
$$
H = - \sum_{l\neq m} \k_\Lambda(a_l - a_m) + 2\eps_1\eps_2\sum_{i=1}^k \ln \left| \frac{\P(\f_i)}{\Lambda^N} \right| + {(\eps_1\eps_2)}^2\sum_{i\neq j} \frac{1}{{(\f_i - \f_j)}^2}
$$
In the thermodynamical limit $k\to\infty$ the number of $\f_i$'s becomes infinite. It is natural to introduce its density. In order to keep the normalizability we define:
\begin{equation}
\label{rhoSU}
\r(x) = \eps_1\eps_2\sum_{i=1}^k \d(x-\f_i).
\end{equation}
In the thermodynamical limit this function becomes smooth. With the help of the density function the Hamiltonian can be rewritten as follows:
$$
H = - \sum_{l\neq m} \k_\Lambda(a_l - a_m) + 2\sum_{l=1}^N\int \dd x\r(x) \ln \left|\frac{x - a_l}{\Lambda}\right| + \vpint_{x\neq y} \dd x\dd y\frac{\r(x)\r(y)}{{(x-y)}^2}.
$$

This expression is rather suggestive. After integration by parts and introducing the {\em profile function}\footnote{in the $SU(N)$ case this function is closely related to the profile of the Young tableaux, written in the Russian style, see \cite{SWandRP} for details.}
\begin{equation}
\label{ProfSU}
f(x) = -2 \r(x) + \sum_{l=1}^N |x-a_l|
\end{equation}
the Hamiltonian can be rewritten in a nice form:
\begin{equation}
\label{HamSUAdjGauge}
H[f] = -\frac{1}{4} \int \dd x\dd y f''(x) f''(y) \k_\Lambda(x-y).
\end{equation}

The partition function \Ref{Z} can be represented as follows:
\begin{equation}
\label{varEqn}
Z(a,m,\eps_1,\eps_2,\Lambda) \sim \int \mathcal{D} f \e^{-\frac{1}{\eps_1\eps_2} H_{\eps_1,\eps_2}[f]}.
\end{equation}
We are interested in the classical approximation of this integral only.

%%% Hamiltonians  %%%

\subsection{Hamiltonians}

%%% Table Hamiltonians %%%

\begin{table}
\begin{center}
%\begin{sideways}
\begin{tabular}{||c|c||c||}
\hhline{|t:=:=:t:=:t|}
\textbf{Group} & \textbf{Multiplet} & \textbf{Contribution to  $H[f]$}\\
\hhline{|:=:=::=:|}	 
& Adjoint, gauge & $ -\dfrac{1}{4}\int \dd x\dd y f''(x)f''(y)\k_\Lambda(x-y)$\\
\hhline{||~|-||-||}
& Fundamental & $  \dfrac{1}{2} \int \dd x f''(x) \k_\Lambda(x+m)$\\
\hhline{||~|-||-||}
$SU(N)$ & Symmetric & $\dfrac{1}{8}\int \dd x\dd y f''(x)f''(y) \k_\Lambda(x+y + m) +  \int \dd x f''(x) \k_\Lambda(x + m/2)$ \\
\hhline{||~|-||-||}
& Antisymmetric & $\dfrac{1}{8} \int \dd x\dd y f''(x)f''(y)\k_\Lambda(x + y + m) -  \int \dd x f''(x) \k_\Lambda(x + m/2)$\\
\hhline{||~|-||-||}
& Adjoint, matter & $\dfrac{1}{4} \int \dd x\dd y f''(x)f''(y)\k_\Lambda(x - y + m)$\\
\hhline{|:=:=::=:|}
& Adjoint, gauge & $-\dfrac{1}{8} \int \dd x\dd y f''(x)f''(y)\k_\Lambda(x + y) + \int \dd x f''(x) \k_\Lambda(x)$\\
\hhline{||~|-||-||}
$SO(N)$ & Fundamental & $  \dfrac{1}{2} \int \dd x f''(x) \k_\Lambda(x+m)$\\
\hhline{||~|-||-||}
& Adjoint, matter & $\dfrac{1}{8} \int \dd x\dd y f''(x)f''(y)\k_\Lambda(x + y + m) -  \int \dd x f''(x) \k_\Lambda(x + m/2)$ \\
\hhline{|:=:=::=:|}
& Adjoint, gauge & $-\dfrac{1}{8} \int \dd x\dd y f''(x)f''(y)\k_\Lambda(x + y) - \int \dd x f''(x)\k_\Lambda(x)$\\
\hhline{||~|-||-||}
$Sp(N)$ & Fundamental & $ \dfrac{1}{2} \int \dd x f''(x) \k_\Lambda(x+m)$\\
\hhline{||~|-||-||}
& Antisymmetric & $\dfrac{1}{8} \int \dd x\dd y f''(x)f''(y)\k_\Lambda(x + y + m)  - \int \dd x f''(x) \k_\Lambda(x + m/2)$\\
\hhline{||~|-||-||}
& Adjoint, matter & $\dfrac{1}{8}\int \dd x\dd y f''(x)f''(y) \k_\Lambda(x+y + m)  + \int \dd x f''(x)\k_\Lambda(x+m/2)$\\ 
\hhline{|b:=:=:b:=:b|}
\end{tabular}
%\end{sideways}
\end{center}
\caption{Hamiltonians}\label{Hams}
\end{table}

%%% end of table %%%

Using the same technique which have lead us to the expression \Ref{HamSUAdjGauge} we can obtain the Hamiltonians for other models listed in the Table \ref{Models}. For orthogonal and symplectic group we use the following definitions of the density and profile functions:

\begin{itemize}
\item \underline{$SO(2n + \c)$}:
\begin{equation}
\begin{aligned}
\label{rhoProfSO}
\r(x) &= \eps_1\eps_2 \sum_{i=1}^k  \Big(\d(x-\f_i) + \d(x+\f_i)\Big), \\
f(x) &= -2\r(x) + \sum_{l=1}^n \Big(|x-a_l| + |x+a_l|\Big) + \c|x|,
\end{aligned}
\end{equation}
\item \underline{$Sp(N)$}: Let $k = 2n + \c$, $\c = 0,1$.
\begin{equation}
\label{rhoProfSp}
\begin{aligned}
\r(x) &= \eps_1\eps_2 \sum_{i=1}^n \Big(\d(x-\f_i) + \d(x+\f_i)\Big), \\
f(x) &= -2\r(x) + \sum_{l=1}^N \Big(|x-a_l| + |x+a_l|\Big).
\end{aligned}
\end{equation}
\end{itemize}
Note that in the case of $SO(N)$ and $Sp(N)$ the density function and the profile function are symmetric.

The Hamiltonians are collected in the Table \ref{Hams}.

%%% Profile function properties %%%

\subsection{Profile function properties}

Let us briefly discuss some properties of the profile function $f(x)$.

First of all we note that since $\r(x)$ has a compact support $f(x)$ behaves like $d |x|$ when $x \to \pm \infty$, where $d$ is the number of connected pieces of the support of $f(x)$. It equals to the dimension of the fundamental representation. 

In general when $|a_l - a_m| \gg \Lambda$ , $l\neq m$, the supporter of $\r(x)$ is a union of $d$ disjoint intervals. Each of them contains one of $a_l$'s. Let $[\a_l^-,\a_l^+]$ be such an interval: $a_l \in [\a_l^+,\a_l^-]$. Then
\begin{equation}
\label{locIntf''}
\int_{\a_l^-}^{\a_l^+} f''(x) \dd x  =  2\int_{\a_l^-}^{\a_l^+} \Big( \d(x - a_l) - \r''(x) \Big) \dd x =2.
\end{equation}
It follows that 
$$
\int_\Real f''(x)\dd x  = 2d.
$$
\begin{equation}
\label{f''(x)xdx}
\begin{aligned}
\int_{\a_l^-}^{\a_l^+} x f''(x) \dd x &= 2\int_{\a_l^-}^{\a_l^+} x \Big( \d(x - a_l) - \r''(x) \Big) \dd x \\
&= 2a_l - 2 {\left. \Big( x\r'(x) - \r(x) \Big)\right|}^{\a_l^+}_{\a_l^-} = 2a_l.
\end{aligned}
\end{equation}
The sum $\sum_{l=1}^d a_l$ equals zero for all group we consider and therefore we have
$$\int_\Real x f''(x) \dd x = 2 \sum_{l=1}^d a_l = 0.$$
Using the definition of  $\r(x)$ for $SU(N)$ \Ref{rhoSU} we have
\begin{equation}
\label{k=int}
\begin{aligned}
\int_\Real x^2 f''(x) \dd x &= 2\int_\Real x^2\left( \sum_{l=1}^N\d(x-a_l) - \r''(x) \right) \dd x \\
&= 2 \sum_{l=1}^N a_l^2 - 4 \int_\Real \r(x)\dd x = 2\sum_{l=1}^N {a_l}^2 - 4\eps_1\eps_2 k.
\end{aligned}
\end{equation}
It follows that this integral fixes the relation between the instanton number $k$ and $\dfrac{1}{\eps_1\eps_2}$.

The equation \Ref{k=int} can be used to represent the factor $q^k$ in the form similar to \Ref{varEqn}. Indeed, we have
\begin{equation}
\label{qasint}
\begin{aligned}
q^k &= \exp-\frac{1}{\eps_1\eps_2}\left\{ - \pi i \t\sum_{l=1}^N a_l^2 + \frac{\pi i \t}{2}\int_\Real x^2 f''(x)\dd x \right\} \\
&= \Lambda^{k\b}\exp-\frac{1}{\eps_1\eps_2}\left\{ - \pi i \t_0\langle a,a\rangle + \frac{\pi i \t_0}{2}\int_\Real x^2 f''(x)\dd x \right\}. 
\end{aligned} 
\end{equation}
The first term in the curly brackets can be identified with the classical prepotential \Ref{PrepClass}. The second term in general should be added to the Hamiltonian. However, for the non-conformal theories, as it was already motioned, $\t_0$ can be neglected, and so this term is irrelevant. It becomes relevant only in the conformal theories.

%%% Lagrange multipliers %%%

\subsection{Lagrange multipliers}

In \Ref{varEqn} the integration is taken only over the functions satisfying the condition \Ref{f''(x)xdx}. This condition is rather complicated to be considered as the definition of the domain of the functional integration.

However we can extend this domain to all the functions after introducing the Lagrange multipliers. The standard way is the following: let $\xi_1,...,\xi_d$ be the multipliers. Then instead of the Hamiltonian $H[f]$ we should minimize the following (Lagrange) functional:
\begin{equation}
\label{Lagrange}
\begin{aligned}
L[f,\xi] &= H[f]  +\sum_{l=1}^d \xi_l \left(  \frac{1}{2}\int_{\a^-_l}^{\a^+_l}x f''(x)\dd x - a_l\right)\\
&= S[f,\xi] - \sum_{l=1}^d \xi_l a_l.
\end{aligned}
\end{equation}
where
\begin{equation}
\label{LegendreF}
S[f,\xi] = H[f]  + \frac{1}{2}\sum_{l=1}^d \xi_l \int_{\a^-_l}^{\a^+_l}x f''(x)\dd x.
\end{equation}
Having found the minimizer $f_\star(x)$ of $L[f,\xi]$ we should also find the stationary point with respect to $\xi_l$. This provide the condition \Ref{f''(x)xdx}. In other words $S[f,\xi]$ should satisfy
\begin{equation}
\label{Legendre}
{\left.\frac{\partial S[f_\star,\xi]}{\partial \xi_l}\right|}_{f_\star = \mr{const}} = a_l.
\end{equation}
where the $\xi$-dependence of  $f_\star(x)$ can be neglected since the derivative of the functional with respect to function is zero at the minimizer. This equation determines $\xi_l$ as some functions of $a_l$. Plugging back these functions into \Ref{Lagrange} we obtain the value of the Hamiltonian at the critical point. That is, the (minus) prepotential. Otherwise the function $S[f_\star,\xi]$ is nothing but the Legendre transform of $-\Prep(a,m)$. 

Note that since $\sum_{l=1}^d a_l = 0$ the sum of $\xi_l$ is not fixed by this procedure.

The last term in \Ref{LegendreF} requires the knowledge of the support of the minimizer $f_\star(x)$ which itself is to be found. Hence the constraints can not be imposed in the form presented above. However another way exists \cite{SWandRP}. Note that $f'(-\infty) = -d$, $f'(+ \infty) = d$ and thanks to \Ref{locIntf''} 
$$f'(\alpha_l^+) - f'(\alpha_l^-) = \int_{\alpha^-_l}^{\alpha^+_l}f''(x)\dd x = 2.$$
Hence we can introduce a piecewise linear function (the {\em surface tension function}) $\sigma(t)$  such that $\sigma'(t) = \xi_l$ when $t = f'(x), x\in [\alpha_l^-,\alpha_l^+]$, that is, $t \in (-d + 2(l-1), -d + 2l )$. With the help of this function we can rewrite the last term in \Ref{LegendreF} as follows
\begin{equation}
\label{Larg}
\frac{1}{2}\sum_{l=1}^d \xi_l \int_{\alpha_l^-}^{\alpha_l^+} x f''(x)\dd x = - \frac{1}{2}\vpint_\Real \sigma(f'(x))\dd x
\end{equation}
provided  $\sigma(d) + \sigma(-d) = 0$. Together with the definition of $\sigma(t)$ it implies $\sum_{l=1}^d \xi_l = 0$ and all the $\xi_l$'s are now defined. 

The discussion presented above implies that in order to determine the prepotential we have proceed the following steps:
\begin{itemize}
\item find the minimizer $f_\star(x)$ of the Lagrange functional:
\begin{equation}
\label{LagrangeS}
S[f,\xi] = H[f] - \frac{1}{2}\vpint_\Real \sigma(f'(x))\dd x,
\end{equation}
where the Hamiltonian $H[f]$ is defined for each model with the help of Table \ref{Hams},
\item in order to obtain the prepotential we need to perform the Legendre transform with respect to $\xi$ of $S[f_\star,\xi]$.
\end{itemize}
As we shall see in the next section the Seiberg-Witten curves appear naturally in these computations.

%%% Seiberg-Witten geometry %%%

\section{Episode III: Back to curves and differentials}
\label{III}

In this section we consider some examples of the saddle point equations. First of all, let us consider an example of $SU(N)$: pure Yang-Mills theory and matter in fundamental representation \cite{SWandRP}.

%%% Example: $SU(N)$, pure Yang-Mills and fundamental matter %%%

\subsection{Example: $SU(N)$, pure Yang-Mills and fundamental matter}
\label{Curve:SU(N)Fund}

Let $N_f$ be the number of flavors. With the help of the Table \ref{Hams} we can write the Hamiltonian of the model:
$$
H[f] = - \frac{1}{4} \int \dd x \dd y f''(x)f''(y)\k_\Lambda(x-y) + \sum_{f=1}^{N_f} \frac{1}{2}\int \dd x f''(x) \k_\Lambda(x+m_f).
$$

In order to minimize the functional \Ref{LagrangeS} we note, that it naturally depends not on $f(x)$, but rather on $f'(x)$. The saddle point (Euler-Lagrange) equation for $f'(x)$ is
\begin{equation}
\label{SPE}
2 \frac{\d S[f,\xi]}{\d f'(x)} =  \int \dd y f''(y) \k'_\Lambda(x-y) - \sum_{f=1}^{N_f}\k'_\Lambda(x+m_f) - \s'(f'(x)) = 0. 
\end{equation}
Using the definition of $\s(t)$ we conclude that $\s'(f'(x)) = \xi_l$ when $x\in [\a_l^-,\a_l^+]$. When $x$ is outside of the support of $f''(x)$, say $x\in(\a_l^+,\a_{l+1}^-)$, we can not determine $\s'(f'(x))$. The only thing we can say is that in this case $\xi_l \leq \s'(f'(x)) \leq \xi_{l+1}$. 

Taking the derivative we obtain:
\begin{equation}
\label{SPE:SU(N)Fund}
\begin{aligned}
\int \dd y f''(y)\ln\left| \frac{x-y}{\Lambda}\right| - \sum_{f=1}^{N_f} \ln \left| \frac{x+m_f}{\Lambda} \right| &= 0, & &x\in[\a_l^-,\a_l^+].
\end{aligned}
\end{equation}

In order to go further we exploit the primitive of the Sokhotski formula:
$$
\ln(x+i0) = \ln|x| - i\pi\Heavi(-x),
$$
where $\Heavi(x)$ is the Heaviside step function: 
$$
\Heavi(x) = \left\{ 
\begin{aligned}
&1, & x &> 0, \\
&0, & x &< 0.
\end{aligned}
\right.
$$
Define the primitive of the resolvent of $f''(x)$:
$$
F(z) = \frac{1}{4\pi i} \int_\Real \dd y f''(y) \ln \left( \frac{z-y}{\Lambda}\right).
$$

%%% Figure SU(N), Fund %%%

\begin{figure}
\includegraphics[width=\textwidth]{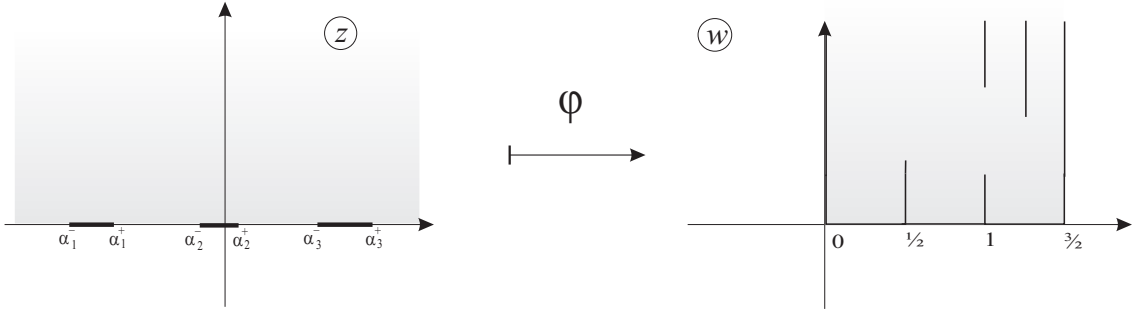}
\caption{Conformal map for $SU(3)$, $N_f=2$}\label{FigSUFund}
\end{figure}

%%% end of figure %%%

For $F(x)$ we obtain the following equation:
\begin{equation}
\label{defOfF}
F(x) - \sum_{a=1}^{N_f}\frac{1}{4\pi i}\ln \left( \frac{x+m_f}{\Lambda}\right) = \vf(x),
\end{equation}
where the complex map $\vf(x)$ maps the real axis to boundary of the domain on the figure \ref{FigSUFund}. It is holomorphic (since the lefthand side is). It follows that $\vf(z)$ maps the upper half-plane to the domain. Suppose that $|a_l - a_m| \gg \Lambda$ if $l\neq m$ and $m_f \gg a_l$ for all $f$ and $l$. This information is sufficient to reconstruct this map. One gets (up to an additive constant):
\begin{equation}
\label{phiForPureSU}
\vf(z) = \frac{1}{2 \pi} \arccos\frac{P(z)}{2 \Lambda^{\b/2} \sqrt{Q(z)}},
\end{equation}
where according to \Ref{ass} $\b = 2N - N_f$ and
$$
\begin{aligned}
Q(x) &= \prod_{f=1}^{N_f} (x + m_f), & P(x) &= \prod_{l=1}^N (x - \a_l).
\end{aligned}
$$
We have introduces parameters $\a_l \in [\a_l^-,\a_l^+]$ which are the classical values of the Higgs vevs.

Define $y(z) = \exp{2 \pi i F(z)}$. Then the solution we have obtained can be written as an equation for $y(z)$:
\begin{equation}
\label{SUfundCurve}
y^2(z) -  P(z)y(z) + \Lambda^\b Q(z) = 0.
\end{equation}
The endpoints of $f''(x)$'s support satisfy the equation
$$
P^2(\a_l^\pm) - \Lambda^\b Q(\a_l^\pm) = 0.
$$

\begin{figure}
\includegraphics[width=\textwidth]{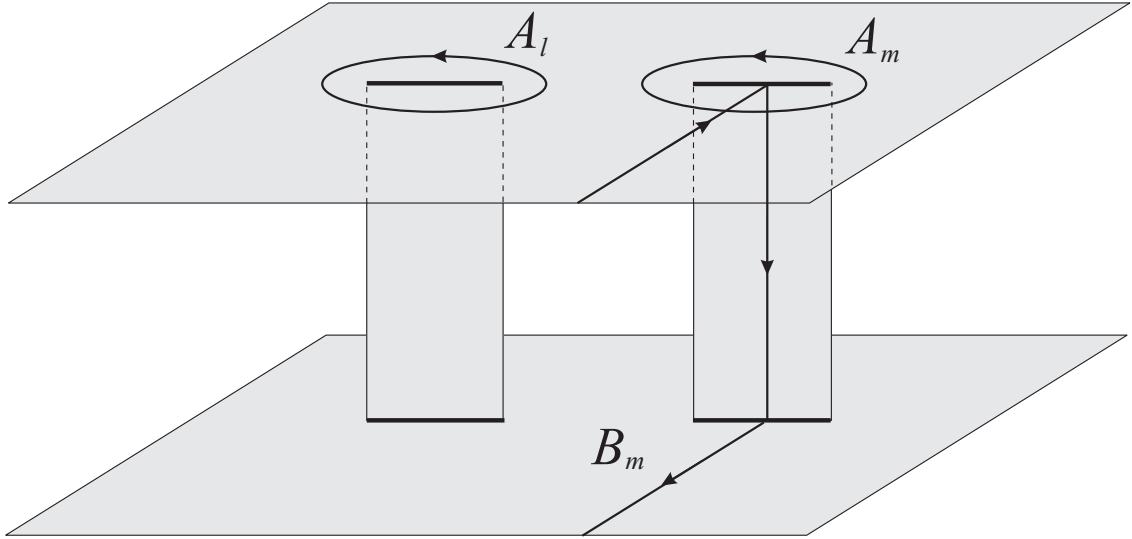}
\caption{Basic cycles}\label{Cycles}
\end{figure}

The Riemann surface of the function $y(z)$ is the two-fold covering of the Riemann sphere. It has cut which connect these two folds along the support of the profile function. Let us define the basic cycles of this Riemann surface (figure \ref{Cycles}). We see that the intersection number satisfies $A_l \# B_m = \d_{l,m}$.

%%% Figure Cycles %%%

%%% end of figure %%%

Using some resolvent properties and \Ref{f''(x)xdx} one shows that
$$
\frac{1}{2} \int_{\a_l^-}^{\a_l^+} x f''(x)\dd x = \oint_{A_l} z \dd F(z) = \oint_{A_l} \frac{1}{2\pi i} z \frac{\dd y}{y} = a_l.
$$

Using the saddle point equation \Ref{SPE} we conclude that
$$
\begin{aligned}
\frac{\xi_{l+1}-\xi_l}{2\pi i} &= 2\int_{\a_l^+}^{\a_{l-1}^-} \left( F(z) - \sum_{f=1}^{N_f} \frac{1}{4\pi i} \ln \left(\frac{z+m_f}{\Lambda}\right)\right) \dd z \\
&= - 2 \int_{\a_l^+}^{\a_{l-1}^-} z \left( \dd F(z) - \sum_{f=1}^{N_f}\frac{1}{4\pi i} \frac{\dd z}{z + m_f} \right) = - \oint_{B_{l+1} - B_l} \frac{1}{2\pi i} z \frac{\dd y}{y}.
\end{aligned}
$$

Performing the Legendre transform inverse to \Ref{Legendre} we obtain
$$
\frac{\pd \Prep}{\pd a_l} = 2\pi i \oint_{B_l} \frac{1}{2\pi i} z \frac{\dd y}{y}.
$$

It follows that the prepotential for this theory can be reconstructed with the help of the Seiberg-Witten data: the curve \Ref{SUfundCurve} and the meromorphic differential 
\begin{equation}
\label{prepSU}
\l = \frac{1}{2\pi i} z \frac{\dd y}{y} =  z \dd F(z).
\end{equation}

%%% Fundamental for SO(N) and Sp(N) %%%

\subsection{Fundamental matter for $SO(N)$ and $Sp(N)$}

In this section we extend the previous analysis to the matter in fundamental representation for other classical groups: $SO(N)$ and $Sp(N)$. The curves for pure Yang-Mills theories have already been discussed in \cite{ABCD}.

%%% SO(N) %%%

\microsection{$SO(N)$ case.}
\label{scn:SOFund}
With the help of the Table \ref{Hams} we obtain the Hamiltonian. In order to obtain the saddle point equation we should take the variation with respect to the {\em symmetric} functions. The function $\s(t)$ is also supposed to be symmetric. The equation we get is
$$
\int \dd y f''(y) \k'_\Lambda(x-y) - 4\k_\Lambda'(x) - \sum_{f=1}^{N_f}\left(\k_\Lambda'(x+m_f) + \k_\Lambda'(x-m_f)\right) - 2 \s'(f'(x)) = 0.
$$
We see that this equation coincides with \Ref{SPE} for $2N_f + 4$ fundamental multiplets with masses $(0,0,0,0,m_1,-m_1,\dots,m_f,-m_f)$. It follows that the same should be true for the prepotential \cite{SOandSp}.

\begin{remark}
One could be warred about $N$-odd case, where one of the Higgs field vevs, which is equals to zero, matches with the zero mass coming form from the term $2\k'_\Lambda(x)$. However already from the expression \Ref{zSOAdjGauge} and \Ref{zSOFund} it is seen that they painless annihilate each other.
\end{remark}

We define $F(z)$, $y(z)$ and $\l$ at the same way as in the $SU(N)$ case. We are able to write the Seiberg-Witten curve (as usual we define $N = 2n + \c$, $\c = 0,1$; according to \Ref{ass} $\b = 2N - 2N_f - 4$):
$$
y^2(z) + z^\c \prod_{l=1}^n(z^2-\a_l^2) y(z) + \Lambda^{\b} z^4 \prod_{f=1}^{N_f}(z^2 - m_f^2) = 0.
$$

%%% Sp(N) %%%

\microsection{$Sp(N)$ case.}
\label{scn:SpFund}

In order to solve the saddle point equation for this model it is convenient to introduce another profile function defined as follows:
\begin{equation}
\label{ProfSpAltern}
\tilde{f}(x) = f(x) + 2 |x| = -2\r(x) + \sum_{l=1}^N\Big(|x-a_l| + |x+a_l|\Big) + 2|x|.
\end{equation}
The new profile function is also symmetric. We also should redefine the surface tension function $\s(t)$ as follows:
$$
\tilde{\s}'(t) = \left\{
\begin{aligned}
-&\xi_l, & t&\in (-2l,-2l-2), & l&=1,\dots,N \\
&0 & t&\in(-2,2) & \\
+&\xi_l, & t&\in (+2l,+2l+2), & l&=1,\dots,N \\
\end{aligned}
\right.
$$

\begin{figure}
\includegraphics[width=\textwidth]{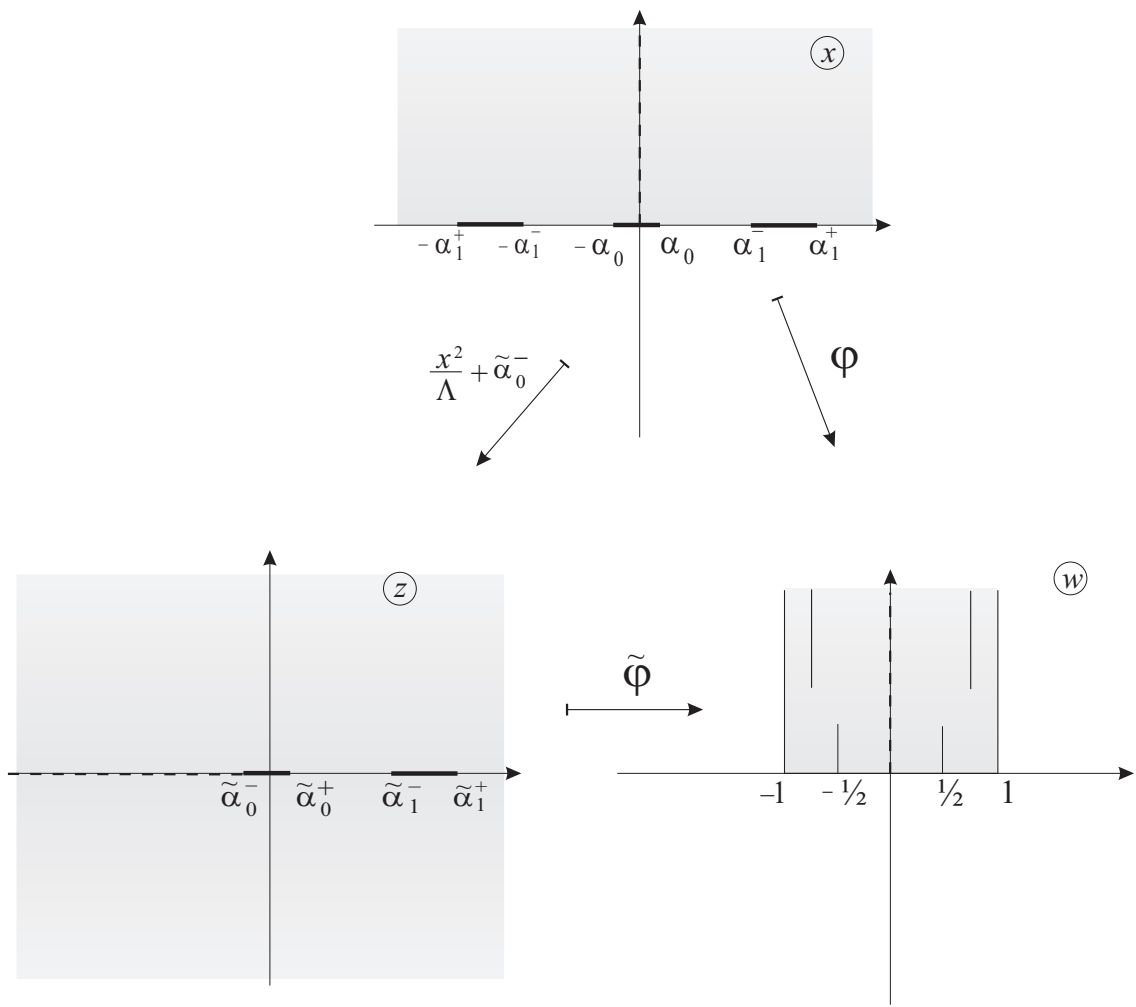}
\caption{Conformal map for $Sp(1)$, $N_f=1$}\label{FigSpFund}
\end{figure}

The Hamiltonian for the gauge multiplet is
$$
\tilde{H}[\tilde{f}] = H[f] = - \frac{1}{8}\int \dd x \dd y \tilde{f}''(x)\tilde{f}''(y) \k_\Lambda(x-y).
$$
And finally the saddle point equation for the model can be written as follows:
$$
\int \dd y \tilde{f}''(y)\k'_\Lambda(x-y) - \sum_{f=1}^{N_f} \left( \k'_\Lambda(x+m_f) + \k'_\Lambda(x-m_f)\right) - 2\tilde{\s}'(\tilde{f}'(x)) = 0.
$$
This equation looks like \Ref{SPE}. However we should remember that the support of $\tilde{f}''(x)$ contains the interval $[\a_o^-,\a_o^+] \ni 0$. Using the definitions \Ref{rhoProfSp} we get
\begin{equation}
\label{SpInt}
\int_{\a_o^-}^{\a_o^+} \tilde{f}''(x) \dd x = 4.
\end{equation}
It follows that for the primitive of the resolvent of $\tilde{f}(x)$ defined by \Ref{defOfF} we obtain the following equation
$$
F(z) - \sum_{f=1}^{N_f} \frac{1}{4\pi i} \left( \ln\left( \frac{z + m_f}{\Lambda} \right) + \ln \left( \frac{z-m_f}{\Lambda} \right) \right) = \vf(z),
$$
where $\vf(z)$ is a holomorphic function which maps the upper halfplain  to the domain on the figure \ref{FigSpFund}. In order to construct the map we use the reflection principle. Function $\vf(z)$ maps first quadrant to the half of the domain. It follows that together with the square map function $\vf(z)$ maps upper halfplain to the half of our domain. Hence we can use the result for $SU(N)$. The map $\tilde{\vf}(z)$ is given by
$$
\begin{aligned}
\tilde{\vf}(z) &= \frac{1}{2\pi} \arccos \frac{z \prod_{l=1}^N(z-\tilde{\a}_l)}{2\Lambda^{N+1-N_f/2} \sqrt{\prod_{f=1}^{N_f} (z + \tilde{m}_f) }}, & \tilde{m}_f = - \tilde{\a}_o^- - \frac{m_f^2}{\Lambda}.
\end{aligned}
$$

%%% Figure Sp(N), Fund %%%

%%% end of figure %%%

The endpoints of the intervals $[\tilde{\a}_l^-,\tilde{\a}_l^+]$ satisfy the equation:
$$
\tilde{\a}_l^\pm \prod_{l=1}^N (\tilde{\a}_l^\pm - \tilde{\a}) = \pm  2 \Lambda^{N+1-N_f/2} \prod_{f=1}^{N_f}\sqrt{\tilde{\a}_l^\pm + \tilde{m}_f}.
$$  
Using this condition we can rewrite the composition of $\tilde{\vf}(z)$ and $z \mapsto z^2 /\Lambda + \tilde{\a}_o^-$ as follows:
$$
\vf(z) = \frac{1}{2\pi}\arccos\frac{z^2\prod_{l=1}^N(z^2 -\a_l^2) + \Lambda^{\b/2} \prod_{f=1}^{N_f}im_f}{2 \Lambda^{\b/2}\sqrt{\prod_{f=1}^{N_f} (z^2 - m_f^2)}},
$$
where $\b = 4N + 4 - 2N_f$.

It follows that the curve can be written as 
$$
y^2(z) + \left[ z^2\prod_{l=1}^N(z^2 -\a_l^2) + \Lambda^{\b/2}\prod_{f=1}^{N_f}im_f \right]y(z) + \Lambda^{\beta} \prod_{f=1}^{N_f} (z^2 - m_f^2) = 0.
$$

%%% Symmetric and antysymmetric reps of SU(N): equal masses %%%

\subsection{Symmetric and antisymmetric representations of $SU(N)$: equal masses}

Another model for which the curve can be obtained with the help of the  analysis of the saddle point equation is the $SU(N)$ gauge theory with symmetric and antisymmetric representations which have equal masses $m$. The table \ref{Hams} shows that the same equation describes the $SU(N)$ gauge theory with two antisymmetric representations with the same masses $m$ and four fundamental multiplets with masses $m/2$. 

Taking into account the discussion after \Ref{qasint} we can write the Hamiltonian of the model as follows
$$
\begin{aligned}
H[f] &= -\frac{1}{4}\int \dd x \dd y f''(x)f''(y) \k_\Lambda(x-y) + \frac{1}{4}\int \dd x \dd y f''(x)f''(y) \k_\Lambda(x+y+m) \\
&- \frac{\pi i \t_0}{2}\int \dd x x^2 f''(x) - \frac{1}{2}\int \dd x \s(f'(x)).
\end{aligned}
$$
The saddle point equation is
$$
\int \dd y f''(y) \k'_\Lambda(x-y) - \int \dd y f''(y) \k'_\Lambda(x+y+m) = 2 \pi i \t_0 x + \s'(f'(x)).
$$ 
Taking the derivative we arrive to
\begin{equation}
\label{SU:SYM+ANT}
\begin{aligned}
\int \dd y f''(y) \ln |x-y| - \int \dd y f''(y) \ln |x+y + m| &= 2 \pi i \t_0,&  x &\in[\a_l^-,\a_l^+]. 
\end{aligned}
\end{equation}

%%% Figure SU(N), Ant + Sym %%%

\begin{figure}
\includegraphics[width=\textwidth]{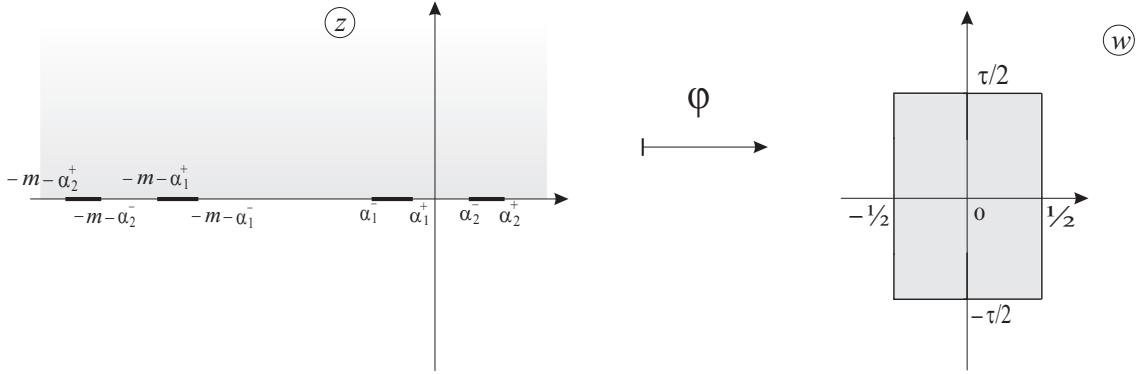}
\caption{Conformal map for $SU(2)$, and matter in symmetric and antisymmetric representation}\label{FigSUAntSym}
\end{figure}

%%% end of figure %%%

The crucial observation is that the function on the lefthand side is antisymmetric under the reflection with respect to $-m/2$: $ x \mapsto - x - m$. So the righthand side is also antisymmetric. Hence the difference of the logarithms equals to $-2i\pi \t_0$ when $x\in[-\a_l^+ - m,-\a_l^- -m]$. Define
\begin{equation}
\label{defFSUSymAnt}
F(z) = \frac{1}{4\pi i} \int_\Real \dd x f''(x) \ln \left( \frac{z-x}{z+x+m}\right).
\end{equation}
The saddle point equation states that $F(z)$ maps the real axis to the boundary of the boundary of the domain on the figure \ref{FigSUAntSym}. So the upper halfplain is mapped to the whole domain. Such a map is known and given by the formula
\begin{equation}
\label{FforSymAnt}
F(z) = \frac{1}{2\o_1 } \sn^{-1}\frac{\vth_3(0)}{\vth_2(0)}\frac{\P(z)}{\P(-z-m)},
\end{equation}
where we have defined $\P(z) = \prod_{l=1}^N (z-\a_l)$. In this formula $\sn(x)$ is the Jacobi elliptic sinus, $2\o_1$ is its real period. It satisfies \cite{Gradstein}
$$
\begin{aligned}
\sn(\o_1 x) &= \frac{\vth_3(0)}{\vth_2(0)}\frac{\vth_1(x)}{\vth_4(x)}, \\
\vth_1(x+1/2) &= \vth_2(x) = \sum_{n\in\mr{odd}} q^{n^2/2} \e^{ i \pi n x}, \\
\vth_4(x+1/2) &= \vth_3(x) = \sum_{n\in\mr{even}} q^{n^2/2} \e^{i \pi n x}.
\end{aligned}
$$
The endpoints of the support of $f''(x)$ satisfy the equation 
$$
\P(\a_l^{\pm}) = \pm \frac{\vth_2(0)}{\vth_3(0)} \P(-m -\a_l^{\pm}).
$$

Using these formulae we can rewrite the expression for $F(z)$ as follows:
$$
\vth_4(2F) \P(z) - \vth_1(2F) \P(-z-m) = 0.
$$

This expression can be checked in various ways. First let us consider the limit $\t_0\to\infty$. In such a limit we have $\vth_3(0)/\vth_2(0) \sim q^{-1/2}$, $\sn(x) \approx \sin(x)$, and $\o_1 \approx \pi$. The expression \Ref{FforSymAnt} becomes
$$
F(z) \approx \frac{1}{2\pi }\sin^{-1} \frac{\P(z)}{q^{1/2}\P(-z-m)}.
$$ 
If we also take a limit $m\to\infty$ in such a way that $m^{2N} q = 4 \Lambda^{2N}$ stays finite we obtain the expression \Ref{phiForPureSU} for pure $SU(N)$ gauge theory, which is consistent with the fact that in this limit the massive representations decouple.

Another way to check this expression is to consider the hyperelliptic truncation of the curve, given by
$$
y(z) + \frac{1}{y(z)} = \frac{\P(z)}{q^{1/2}\P(-z-m)},
$$
where $y(z) = i \exp{2\pi i F(z)}$. Comparing this expression with \Ref{HyperEllipticTrunc} and referring to the Table \ref{z1rules} we see that the one instanton corrections are correctly described by this curve.

%%% Mapping to SU(N) case%%%

\subsection{Mapping to $SU(N)$ case}
\label{MappingToSU}

As anther application, the saddle point equations help to establish the connection between different models. Some of them have been already found after examination the Seiberg-Witten curve \cite{SOandSp}, and the 1-instanton corrections \cite{EllipticMod}. In this section we will examine the saddle point equation. If for two theories they match (after the appropriate identification the parameters of curves) it is natural to expect that the prepotentials will be the same.

As an example consider the $SU(N)$ theory with the symmetric and antisymmetric matter and some fundamental matter. We have the following saddle point equations:

\microsection{Antisymmetric matter.}
\begin{multline}
\label{SPE:SU(N)ANT}
\int \dd y f''(y) \logmodL {x-y} - \frac{1}{2} \int \dd y f''(y)\logmodL{x+y+m^{(a)}} + 2\logmodL{x+m^{(a)}/2} \\
-\sum_{f=1}^{N_f^{(a)}} \logmodL{x+m_f} = 2\pi i \t_0 + \s'(f'(x)), \;\;\;\;\; x\in[\a_l^-,\a_l^+]. 
\end{multline}
\microsection{Symmetric matter}
\begin{multline*}
\int \dd y f''(y) \logmodL {x-y} - \frac{1}{2} \int \dd y f''(y)\logmodL{x+y+m^{(s)}} - 2\logmodL{x+m^{(s)}/2} \\
-\sum_{f=1}^{N_f^{(s)}} \logmodL{x+m_f} = 2\pi i \t_0 + \s'(f'(x)), \;\;\;\;\; x\in[\a_l^-,\a_l^+]. 
\end{multline*}
The analysis of these two expressions leads us to the conclusion that the matter in the symmetric representation with mass $m$ is equivalent to the matter in antisymmetric representation with the same mass together with four fundamental multiplets with masses $m/2$.

In the Table \ref{MappingToSU} we list such equivalences between the models containing different groups and matter content. For each model we find its $SU(N)$ partner. We use the following notation: $\vec{a} = (a_1,\dots,a_n)$ for $SO(2n+\c)$ models and $\vec{a} = (a_1,\dots,a_N)$ for $Sp(N)$ models. As usual, for the orthogonal group $SO(N)$ notation $\diamondsuit$ means 0 when $N$ is odd and it is absent when $N$ is even.

%%% Table Mapping to SU(N) %%%

\begin{table}
\begin{center}
\begin{tabular}{||c|c||c|c|c||}
\hhline{|t:=:=:t:=:=:=:t|}
\textbf{Group} & \textbf{Multiplet} & \textbf{Higgs} & \textbf{Fund.} & \textbf{Anti.} \\
\hhline{|:=:=::=:=:=:|}
$SU(N)$ & Symmetric, $m$ & $\vec{a}$ &  $m/2$, $m/2$, $m/2$, $m/2$ & $m$ \\
\hhline{|:=:=::=:=:=:|}
& Adjoint, gauge & $(\diamondsuit,\vec{a},-\vec{a})$ &0, 0, 0, 0 & --- \\
\hhline{||~|-||-|-|-||}
$SO(N)$ & Fundamental, $m$ & $(\diamondsuit,\vec{a},-\vec{a})$ & $-m$, $+m$  & --- \\ 
\hhline{||~|-||-|-|-||}
& Adjoint, $m$ & $(\diamondsuit,\vec{a},-\vec{a})$ & --- & $+m$, $-m$ \\
\hhline{|:=:=::=:=:=:|}
& Adjoint, gauge & $(0,0,\vec{a},-\vec{a})$ & --- &  --- \\
\hhline{||~|-||-|-|-||}
& Adjoint, gauge &  &  & \\
$Sp(N)$& + 2 fund., $m=0$ & $(\vec{a},-\vec{a})$ & --- & --- \\
\hhline{||~|-||-|-|-||}
& Fundamental, $m$ & $(\vec{a},-\vec{a})$ & $+m$, $-m$ & --- \\ 
\hhline{||~|-||-|-|-||}
& Antisymmetric, $m$ & $(\vec{a},-\vec{a})$ & --- & $+m$, $-m$ \\
\hhline{||~|-||-|-|-||}
& Adjoint, $m$  &$(\vec{a},-\vec{a})$  & $+m/2$,$+m/2$,$-m/2$,$-m/2$ & $+m$, $-m$  \\
\hhline{|b:=:=:b:=:=:=:b|}
\end{tabular}
\end{center}
\caption{Mapping to $SU(N)$}\label{ToSU}
\end{table}

%%% end of table

%%% Hyperelliptic approximations %%%

\subsection{Hyperelliptic approximations}
\label{HyperEllipticApproximation}

In this section we show how to extract the hyperelliptic approximation to the Seiberg-Witten curve from the saddle point equation. This allows us to prove that the 1-instanton corrections which will be obtained from the curves match with our computation presented in section \ref{1instComp} 

In \ref{1instComp} we have shown that our computations agree with the algebraic curve computation provided the curve is given by \Ref{HyperEllipticTrunc} and the residue function have been constructed with the help of the Table \ref{Srules}. It follows that the only thing we should show is that when solving the saddle point equation in hyperelliptic approximation we obtain the correct rules for the residue function.

Note that for all (classical) groups and fundamental matter the hyperelliptic approximation is exact. It follows that the task is already accomplished for these models. 

Consider the first non-trivial case, the antisymmetric representation for $SU(N)$ model.

%%% SU(N), antisymmetric matter and some fundamentals %%%

\microsection{$SU(N)$, antisymmetric matter and some fundamentals.} The saddle point equation for this model is given by \Ref{SPE:SU(N)ANT}. In order to obtain the hyperelliptic approximation to the Seiberg-Witten curve we will simplify the second term. 

To do that we note that the approximation to the profile function which leads to the perturbative prepotential is the following (see \Ref{ProfSU}):
$$
f_{\mr{pert}}(x) = \sum_{l=1}^N |x-a_l|.
$$
The second derivative of this function has a pointwise support. The support of the exact solution is the union of intervals which has length of order $\Lambda \ll m$. Consider the primitive of the resolvent of $f''(x)$:
$$
F(z) = \frac{1}{4\pi i} \int_\Real \dd y f''(y) \ln \left(\frac{z-y}{\Lambda}\right).
$$
The primitive of $f_{\mr{pert}}$-resolvent is
$$
F_{\mr{pert}}(z) = \frac{1}{2\pi i }\sum_{l=1}^N \ln \left( \frac{z-a_l}{\Lambda}\right).
$$
The exact expression for $F(z)$ will be different. However, if $|z-a_l| \gg \Lambda$ for all $l=1,\dots,N$ we can still use this approximation. In particular when we compute integral over the cycles $A_l$ or $B_{l+1} - B_{l}$ we can use for $F(-z-m)$ the perturbative approximation. Coming back to the equation \Ref{SPE:SU(N)ANT} we conclude that in order to obtain 1-instanton correction we can put in the second term $f(x) = f_{\mr{pert}}(x)$. After this identification the equation becomes:
\begin{multline*}
\int \dd y f''(y) \logmodL{x-y} - \sum_{l=1}^N  \logmodL{x+a_l+m} \\
+ 2\logmodL{x+m/2} - \sum_{f=1}^{N_f} \logmodL{x+m_f} = 2\pi i \t_0, \;\;\;\;\; x\in [\a_l^-,\a_l^+].
\end{multline*}
To solve this equation let us define another profile function
$$
\tilde{f}(x) = f(x) + |x+m/2|.
$$
For this function we have the following saddle point equation:
%\begin{multline*}
$$
\begin{aligned}
&\int \dd y \tilde{f}''(y) \logmodL{x-y} - \sum_{l=1}^N  \logmodL{x+a_l+m} - \sum_{f=1}^{N_f} \logmodL{x+m_f} = 2\pi i \t_0, & x&\in [\a_l^-,\a_l^+].
\end{aligned}
$$
%\end{multline*}
This equation looks like \Ref{SPE:SU(N)Fund} if we identify $\vec{a}\mapsto (-m/2,\vec{a})$, and $m_f \mapsto (-m-a_1,\dots,-m-a_N,m_1,\dots,m_{N_f})$. Using the result of section \ref{Curve:SU(N)Fund} we can immediately write the solution \Ref{SUfundCurve}:
\begin{equation}
\label{HyperSU(N)ANT}
y(z) + \frac{1}{y(z)} = \frac{(2z + m)\prod_{l=1}^N (z-\a_l)}{\Lambda^{(N+2-N_f)/2}\sqrt{\prod_{l=1}^N(z+m+a_l)\prod_{f=1}^{N_f}(z+m_f)}}.
\end{equation}
\begin{remark}
Since we have identified the mass of the antisymmetric multiplet with one the Higgs vevs we should, in principle, write its contribution to the nominator as $(2z + \mu)$, where $\mu = m + O(\Lambda^{\b/2})$. However in order to compute the prepotential we will not need to compute any contour integral where contour passes near the point $-m/2$. It follows that the shift $\mu\mapsto m$ will take effect only in the higher instanton corrections which we are not interested in here.
\end{remark}

The equation \Ref{HyperSU(N)ANT} is the same as \Ref{HyperEllipticTrunc} provided we set$$
S(x) = \frac{\prod_{l=1}^N (x+m+a_l)\prod_{f=1}^{N_f}(x+m_f)}{{(2x+m)}^2\prod_{l=1}^{N_f}{(x-\a_l)}^2}.
$$
This expression matches with the value of the residue function which can be build with the help of the Table \ref{Srules}. The last observation proves that the solution of the saddle point equation \Ref{SPE:SU(N)ANT} gives the correct prediction for the 1-instanton correction.

The procedure presented above can be easily converted to the mnemonic rule to build the residue function which appears in \Ref{HyperEllipticTrunc}. It can be formulated as follows: any term of the form 
$$
\ep \logmodL{x-x_0}
$$ 
leads  to the ${(x-x_0)}^{-\ep}$ factor of the $S(x)$.

%%% SU(N), matter in the symmetric representation %%%

\microsection{$SU(N)$, matter in the symmetric representation.} In order to obtain the hyperelliptic approximation for the case of symmetric representation we can either use the same technique as in the case of the antisymmetric multiplet or directly apply the result of the section \ref{MappingToSU}. Anyway the result for the simplified saddle point equation is
$$
\begin{aligned}
&\int \dd y f''(y) \logmodL{x-y} - \sum_{l=1}^N  \logmodL{x+a_l+m} -2 \logmodL{x+m/2}= 2\pi i \t_0, &  x&\in [\a_l^-,\a_l^+].
\end{aligned}
$$
Applying our rule we get the following contribution to the residue function:
$$
{(2z+m)}^2 \prod_{l=1}^N (z+m+a_l)
$$
which is in the agreement with the Table \ref{Srules}.

%%% SU(N), matter in the adjoint representation %%%

\microsection{$SU(N)$, matter in the adjoint representation.} The contribution to the simplified saddle point equation is
$$
-\sum_{l=1}^N \logmodL{x - a_l + m}.
$$
It follows that the contribution to the residue function is
$$
\frac{1}{\prod_{l=1}^N (x-a_l + m)}.
$$
It agrees with the Table \ref{Srules}.

%%% SO(N) models %%%

\microsection{$SO(N)$ models.} In order to establish the same results for the $SO(N)$ models we can apply the result of the section \ref{MappingToSU}. The result for the adjoint gauge multiplet is:
$$
S(x) = \frac{x^{4-2\c}}{\prod_{l=1}^N{(x-\a_l)}^2}.
$$
For the adjoint matter multiplet we get the following contribution to the residue function:
$$
\frac{{(x^2 - m^2)}^\c}{4x^2 - m^2}\prod_{l=1}^N ({(x + m)}^2 - a_l^2)({(x - m)}^2 - a_l^2).
$$
These expression are in agreement with the Table \ref{Srules}.

%%% Sp(N) models %%%

\microsection{$Sp(N)$ models.} Using the result of section \ref{MappingToSU} we get the following residue function for the gauge multiplet:
$$
S(x) = \frac{1}{x^4\prod_{l=1}^N {(x^2 - \a_l^2)}^2}.
$$
The factors which come from the antisymmetric representation is defined by the following contribution to the saddle point equation
\begin{multline*}
-\frac{1}{2}\int \dd y f''(y) \logmodL{x+y+m}-\frac{1}{2}\int \dd y f''(y) \logmodL{x+y-m} \\
+ 2\logmodL{x+m} + 2 \logmodL{x-m} + 2\logmodL{x+m/2} + 2 \logmodL{x-m/2}.
\end{multline*}
Plugging into this expression the perturbative approximation of the profile function
$$
f_{\mr{pert}}(x) = \sum_{l=1}^N (|x-a_l| + |x+a_l|) + 2|x|
$$
we obtain the following contribution to the residue function:
$$
\frac{\prod_{l=1}^N {({(x+m)}^2 - a_l^2)({(x-m)}^2 - a_l^2)}^2}{(4x^2 - m^2)}.
$$
The contribution to the residue function which comes from the adjoint representation is
$$
(4x^2 - m^2)\prod_{l=1}^N {({(x+m)}^2 - a_l^2)({(x-m)}^2 - a_l^2)}^2.
$$
Obtained expressions is in agreement with the Table \ref{Srules}.

%%% Discussion and further directions %%%

\section{Discussion and further directions}

In this paper have derived the method which allows us to compute the low-energy effective action for $\N = 2$ supersymmetric Yang-Mills theories. We have considered {\em all} the models allowed by the asymptotic freedom. Using the results of \cite{EnnesSU2AntFund,EnnesMasterFunc,EllipticMod,MtheoryTested} we have shown in section \ref{1instComp} that the equivariant deformation method provides the results which in the 1-instanton level agree with the previous computations. 

Also we have written the saddle point equation for each models and we have shown that in all cases when we can it solve obtained expressions for the Seiberg-Witten data agree with known results. 

We have shown that the saddle point equation technique is self-consistent: in spite of the fact that the curves and the differentials are obtained under rather strong condition $k\to\infty$, the final answer is nevertheless correct even if $k$ is low.

In section \ref{HyperEllipticApproximation} we obtained the hyperelliptic approximation to the Seiberg-Witten curve. Presumably, one can develop the method presented there and obtain the mathematically rigorous recursion scheme which will give all the instanton corrections. It would be interesting to establish its relation with other recursion schemes (such as, for example, \cite{Liouville}).   

Another direction would be the generalization of the moduli space singularities counting. Close relation between these singularities and Young tableaux allows us to compute the integral \Ref{partition} (see \cite{SWfromInst}). It would be interesting to generalize this approach to other models.

\section{Acknowledgments} The author thanks Nikita Nekrasov for numerous helpful discussion. Without his constant support it would have not been possible to come into the subject.

%%% References %%%

%\bibliographystyle{amsplain}
%\bibliography{d:/Articles/Catalog/Data/BibTeX/Articles}

\providecommand{\bysame}{\leavevmode\hbox to3em{\hrulefill}\thinspace}
\providecommand{\MR}{\relax\ifhmode\unskip\space\fi MR }
\providecommand{\MRhref}[2]{%
  \href{http://www.ams.org/mathscinet-getitem?mr=#1}{#2}
}
\providecommand{\href}[2]{#2}

\end{document}